\documentclass[useAMS,usenatbib,usegraphicx,usefloat]{mn2e}
\def\fp1{\mbox{$\log(R_e)= a\ \log(\sigma) + b\ \langle \mu \rangle_e +c$}}
\def\muem{\mbox{$\langle \mu \rangle_{\rm e}$}}
\def\re{\mbox{$R_{\rm e}$}}
  
\def\H0{\mbox{$H_0$}}
\def\q0{\mbox{$q_0$}}

\def\rnn{\mbox{$r^{1/n}$}}
\def\kms{\rm km~s$^{-1}$}

\def\eg{{\it e.g.\/}}
\def\ie{{\it i.e.\/}}

\title[The hybrid solution for the FP]{The hybrid solution for the Fundamental Plane}
\author[M. D'Onofrio et al.]{M. D'Onofrio$^{1}$\thanks{E-mail:mauro.donofrio@unipd.it}, G. Fasano$^{2}$ , A. Moretti$^{1}$,  P. Marziani$^{2}$,  D. Bindoni$^{1}$,  J. Fritz$^{3}$,   
\newauthor J. Varela$^{7}$, D. Bettoni$^{2}$, A. Cava$^{4}$, B. Poggianti$^{2}$, M. Gullieuszik$^{2}$,
\newauthor P. Kj{\ae}rgaard$^{6}$, M. Moles$^{7}$, B. Vulcani$^{8}$, A. Omizzolo$^{9,2}$, W.J. Couch$^{10}$, A. Dressler$^{11}$\bigskip\noindent \\
$^1$Astronomy Department, Vicolo Osservatorio 3, I-35122 Padova, Italy \\
$^2$INAF/Astronomical Observatory of Padova, Vicolo Osservatorio 5, I-35122 Padova, Italy \\
$^3$Sterrenkundig Observatorium, University of Gent Krijgslaan 281 S9, B-9000 Gent, Belgium \\
$^4$Astrophysics Department of the University Complutense of Madrid, 28040 Madrid, Spain \\
$^6$The Niels Bohr Institute for Astronomy Physics and Geophysics, Juliane Maries Vej 30, 2100, Copenhagen, Denmark \\
$^7$Centro de Estudios de Fisica del Cosmos de Arag\'on, Plaza de San Juan 1, 44001 Teruel, Spain \\
$^8$Kavli Institute for the Physics and Mathematics of the Universe, University of Tokyo, Kashiwa, 277-8583, Japan \\
$^9$Vatican Observatory Research Group, University of Arizona, Tucson, AZ 85721, USA \\
$^{10}$Center for Astrophysics and Supercomputing, Swinburne University of Technology, PO Box 218, Hawthorn Victoria 3122, Australia \\
$^{11}$Observatories of the Carnegie Institution of Washington, Pasadena, CA 91101, USA}

\begin{document}

\date{}

\pagerange{\pageref{firstpage}--\pageref{lastpage}} \pubyear{2012}

\maketitle

\label{firstpage}

\begin{abstract}

  By exploiting the database of early-type galaxies (ETGs) members of
  the $WINGS$ survey of nearby clusters, we address here the long
  debated question of the origin and shape of the Fundamental Plane
  (FP). Our data suggest that different physical mechanisms concur in
  shaping and {\it 'tilting'} the FP with respect to the virial plane (VP)
  expectation. In particular, an ``hybrid solution'' in which the structure 
  of galaxies and their stellar population are the main contributors to the
  FP {\it tilt} seems to be favoured.
  
  We find that the bulk of the {\it tilt} should be attributed to
  structural non-homology, while stellar population effects play an
  important but less crucial role. In addition, our data indicate
  that the differential FP {\it tilt} between the $V$- and $K$-band is
  due to a sort of entanglement between structural and stellar
  population effects, for which the inward steepening of color
  profiles ($V-K$) tends to increase at increasing the stellar mass of
  ETGs.

  The same kind of analysis applied to the $ATLAS3D$ and $SDSS$ data
  in common with $WINGS$ ($WSDSS$ throughout the paper) confirms
  our results, the only remarkable difference being the less important
  role that our data attribute to the stellar mass-to-light-ratio
  (stellar populations) in determining the FP {\it tilt}. The $ATLAS3D$
  data also suggest that the FP {\it tilt} depends as well on the dark
  matter (DM) fraction and on the rotational contribution to the
  kinetic energy ($V_{rot}/\sigma$), thus again pointing towards the
  above mentioned ``hybrid solution''.

  We show that the global properties of the FP, \ie\ its {\it tilt}
  and tightness, can be understood in terms of the underlying
  correlation among mass, structure and stellar population of ETGs,
  for which, at increasing the stellar mass, ETGs become (on average)
  'older' and more centrally concentrated.

  Finally, we show that a Malmquist-like selection effect may mimic a
  differential evolution of the mass-to-light ratio for galaxies of
  different masses. This should be taken into account in the studies
  investigating the amount of the so called ``downsizing'' phenomenon.

\end{abstract}


\section{Introduction}\label{Intro}

Since its discovery, due to \citet{DjorgDavis} and \citet{Dressetal},
the Fundamental Plane (FP), \ie\ the relation linking the effective
radius (\re), the central velocity dispersion ($\sigma$) and the
average effective surface brightness (\muem) of early-type galaxies
(ETGs):

\begin{equation}\label{fp}
\fp1
\end{equation}

\noindent
has been considered a key tool for investigating the physical
mechanisms driving their formation and evolution. In fact, the
observed {\it tilt} of the FP with respect to the virial plane (VP)
and its tightness imply a peculiar connection between the structure of
the galaxies, their history of star formation and their dark matter
(DM) content, offering useful constraints for the theoretical models.

The {\it tilt} problem arose from the observation that, under the
assumption of homology and constant $M/L$, the FP coefficients ($a$,
$b$) deviate significantly from the virial expectation ($a=2$,
$b=0.4$).  The typical observed values are in fact $a\sim1.2$ and
$b\sim0.3$ in the $V$ band, with small variations mainly depending on
the adopted waveband \citep[see \eg][]{Scodeggio,LaBarbera}.

Since ETGs were considered for a long time ``homologous'' stellar
systems, the FP {\it tilt} was originally attributed to stellar
population effects, on the basis of the argument that its very
existence, under virial equilibrium conditions (and assuming
homology) implies a correlation between the dynamical mass-to-light
ratio and the galaxy mass: $M/L \sim M^{\alpha}$. \citet{Faber87}
found $\alpha\sim0.25$ and later, independent analyses found similar
values of $\alpha$ \citep[see \eg ][]{Pahre,Gerhard,Borriello,Treu05}.

Among the factors potentially causing the variation of the
mass-to-light ratio along the FP, the stellar metallicity, age and
initial mass function (IMF), as well as the DM fraction were first
considered.  The trend in the mean metallicity seemed a viable option
\citep[see \eg][]{Gerhard}, but its effect was estimated to produce
only a small fraction of the {\it tilt} \citep[see \eg\
][]{DjorgDavis,DjorgSant}.  Stellar population synthesis models failed
to reproduce the {\it tilt} \citep{Renzini1995}, but \cite{Chiosi} and
\cite{ChiosiCarr}, in the context of a ``monolithic'' scenario of
galaxy formation, were able to explain it using a varying IMF and a
different SFH for galaxies of different masses.  The existence of a
variable IMF is now supported by several papers \citep[see \eg\ for a
review][]{Kroupa2,Cappellari2}.  However, a drastic IMF variation
seems required to produce the observed {\it tilt} \citep[see
\eg][]{Renzini}, while a SFH that smoothly varies with mass is more
difficult to reconcile with the widely accepted ``hierarchical''
merging paradigm of the $\Lambda$CDM cosmology.  A significant
contribution of the DM to the FP {\it tilt} was excluded by
\citet[][hereafter, C96]{Ciotti} on the basis of a fine-tuning
argument, but \cite{Tortora}, estimating the total $M/L$ ratio from
simple Jeans dynamical models, found that the DM fraction within the
effective radius \re\ is roughly constant for galaxies fainter than
$M_B \sim -20.5$ and turns out to increase for brighter galaxies, thus
implying a systematic variation of the dark-to-bright matter ratio
along the FP.  \cite{Padmanabhan} and \cite{Hyde} also found evidence
that the DM fraction ($M_{tot}/M^*$) increases with mass.

The finding by \cite{Pahre} that the {\it tilt} is still substantial
in the $K$-band, where the luminosity maps the bulk of stellar mass,
prompted the search for new explanations of its origin, not directly
related to stellar population effects.  Many works proposed the
alternative scenario in which the ``broken structural and dynamical
homology'' of ETGs is at the origin of the {\it tilt} \citep{Hjorth,
  PrugSimien, Busarello, GrahamColless, Pahre, Bertin, Trujillo,
  Nipoti, LaBarbera}. This interpretation was supported by the
observation that ETGs are ``non-homologous'' stellar systems in both
their structure and dynamics \citep{Capaccioli87, deCarvalho88,
  Capaccioli89, Burkert93, Michard85, Schombert86, Caon, YoungCurrie,
  PrugSimien}. However, C96 claimed that again a strong fine--tuning
between stellar mass-to-light ratio and structure (Sersic index $n$)
is required to explain with just structural non-homology both the {\it
  tilt} of the FP and the small scatter around it. The role of
non-homology was also excluded by \cite{Cappellari} and
\cite{Cappellari3} using integral models of the ETGs mass distribution
based on 2D kinematic maps. Along the same vein, \cite{Bolton}, using
the galaxies masses estimated from the gravitational lensing, claimed
that structural non-homology does not have a significant role in {\it
  tilting} the FP.

The tightness of the FP relation is particularly important because
it provides the strongest constraints on the SFH of galaxies.
The origin of the FP scatter was investigated by \cite{Forbes} and
\cite{Terlevich}, who found a correlation between the residuals of the
FP and the age of the galaxies (ETGs with higher/lower surface
brightness have younger/older ages).  \cite{Gargiulo} found that the
FP residuals anti-correlate with the mean stellar Age, while a strong
correlation exists with $\alpha/Fe$.  In this case, the distribution
of galaxies around the FP is tightly related to enrichment, and
hence to the timescale of star-formation.  \cite{Gravesetal} found
that the stellar population variations contribute at most 50\% of the
total thickness and that correlated variations in the IMF or in the
central DM fraction make up the rest.  Recently, \cite{Magoulas},
using a sample of $10^4$ ETGs extracted from the 6dF Galaxy Survey,
found that the residuals about the FP show significant trends with
environment, morphology and stellar population, the strongest
trend being with age.

This short review of the FP problem makes it clear that a general
consensus about the origin of its properties is still lacking. In
particular the role played by non-homology is far from being fully
understood.

In \citet[Paper-I]{Donofrio1}, we studied the FP of a sample of 1550
ETGs, obtained cross-matching the $V$-band, surface photometry dataset
of ETGs from $WINGS$ \citep[Wide-field Imaging of Nearby Galaxy-clusters
Survey]{Fasano,Varela} with velocity dispersions from literature data
\citep{Smith,Bernardi}. Our main conclusions were the
following: 1) the FP coefficients depend on the luminosity range of
the ETGs sample under analysis, as well as on the fitting strategy; 2)
the FP coefficients do depend on the local density, while they do not
depend on the global cluster properties (such as \eg\ X-ray emission);
3) the stellar mass-to-light ratio ($M^*/L$) does not correlate with
the $V$-band luminosity, so that a possible role of non-homology in
causing the FP {\it tilt} should be considered.

In this paper we exploit the spectroscopic and photometric, $K$-band
database of ETGs in the $WINGS$ survey \citep{Fritz1,Valentinuzzi} to
complete the analysis of the FP problem.

The paper is organized as follows: in Sec.~\ref{sec1} we present the
main equations defining the FP problem. In Sec.~\ref{sec2} we describe
the $WINGS$ data samples used in this work.  In Sec.~\ref{sec3} we
derive the FP coefficients for the $V$- and $K$-band.  In
Sec.~\ref{sec4} we discuss the origin of the bulk of the FP {\it tilt}
and compare our results with those obtained using the $ATLAS3D$ and
$WSDSS$ databases. In Sec.~\ref{sec6} we investigate the origin of the
differential {\it tilt} observed between the $V$ and $K$ wavebands. In
Sec.~\ref{sec7} we address the problem of the thickness of the FP and
its connection with the non-homology of ETGs through the
structure--stellar~population conspiracy. In Sec.~\ref{sec8} we
discuss the variation of the $M/L-M$ relation with redshift and the
occurrence of selection effects. Our conclusions are summarized in
Sec.~\ref{sec9}, where we also try to probe our findings against the
present theoretical models of galaxy formation and evolution.

In this paper we use $H_0=70$ km sec$^{-1}$ Mpc$^{-1}$,
$\Omega_{\Lambda}=0.7$ and $\Omega_m=0.3$.

\section{The FP problem in a nutshell}\label{sec1}

We assume that ETGs are gravitationally bound stellar systems which
satisfy the virial theorem equation:

\begin{center}
\begin{equation}\label{eqvir}
\langle V^2 \rangle \propto \frac{GM_{tot}}{\langle R \rangle},
\end{equation}
\end{center}

\noindent
where $M_{tot}$ is the total galaxy mass, $\langle R \rangle$ is a
proxy for the gravitational radius, and $\langle V^2 \rangle$ the mean
kinetic energy per unit mass. Ideally, all virialized systems should
be placed onto the VP in the space defined by the
variables $M_{tot}$, $\langle R \rangle$ and $\langle V^2 \rangle$.
Unfortunately, these are not observable quantities. Therefore, in the
case of ETGs, the virial equation ~(\ref{eqvir}) is usually written as
follows:

\begin{center}
\begin{equation}\label{eqMtot}
M_{tot} \propto \frac{K_V\sigma^2\re}{G},
\end{equation}
\end{center}

\noindent
where $\sigma$ is the central velocity dispersion within a fixed
aperture, \re\ is the equivalent radius of the isophote enclosing half
the total galaxy luminosity and $K_V=1/(k_vk_r)$ takes into account
projection effects, density distribution and stellar orbits
distribution. The term $K_V$ parametrizes our ignorance about
orientation, 3D structure and dynamics of ETGs.  The formal expression
of $K_V$ assumes $\langle V^2 \rangle = k_v \sigma^2$ and $\langle R
\rangle = k_r\re$.

Introducing the mean effective surface brightness $\langle I \rangle_e
= L/2\pi \re^2$, one gets:

\begin{center}
\begin{equation}\label{eqvir5}
\re \propto \frac{K_V}{2\pi G}\ (\frac{M_{tot}}{L})^{-1}\ \langle I\rangle_e^{-1}\ \sigma^2,
\end{equation}
\end{center}

\noindent
or, in log units and after some algebra:

\begin{center}
\begin{equation}\label{eqvirlog}
\log(\re) = 2\log(\sigma)+0.4\muem +C_\lambda +
\end{equation}
\[
\ \ \ \ \ \ \ \ \ \ \ \ +\log(K^*_V)-\log(\frac{M^*}{L}),
\]
\end{center}

\noindent
where $C_\lambda = - 0.4[M_\odot(\lambda) + 21.572] -\log(2\pi G)$ and
$M_\odot(\lambda)$ is the absolute magnitude of the Sun in the given
band.  In the equation~(\ref{eqvirlog}), we have replaced the total
mass $M_{tot}$ with $M^*$ (stellar mass) and $K_V$ with
$K^*_V=K_V/k_m=1/(k_vk_rk_m)$, where $k_m=M_{tot}/M^*$ parametrizes
our ignorance about the DM content.

This formulation of the Virial theorem is directly comparable with the
FP equation (\ref{fp}) empirically derived from observations. It
formally illustrates the problem of the FP {\it tilt}, given that, in
all photometric bands, the observed coefficients of $\log(\sigma)$ and
$\muem$ turn out to be remarkably different from the
virial expectation. Since we assume that ETGs are in virial equilibrium, the
reason for the observed deviation of the FP coefficients from the
virial expectation must reside in the term:

\begin{equation}\label{eqKV1}
\Delta_{FP}=\log(K^*_V)-\log(\frac{M^*}{L}).
\end{equation}

\noindent
In fact, any systematic dependence of this expression on the position
along the FP would produce a displacement ({\it tilt}) of the FP from
the VP.

According to eq.~(\ref{eqKV1}), besides the stellar populations
($M^*/L$), the structural and dynamical non-homology ($k_r$ and $k_v$,
respectively), as well as the dark matter fraction ($k_m$) might
contribute to the FP {\it tilt}.  The relative importance of each
factor can be in principle estimated by determining how strongly it
correlates with the position along the FP.

In order to parametrize such position, we note that, since each galaxy
must simultaneously lie onto the FP and on its proper VP, the right
terms of equations (\ref{fp}) and (\ref{eqvirlog}) can be equated,
thus giving:

\begin{equation}\label{eqvirobs}
(a-2)\log(\sigma)+(b-0.4)\muem+c-C_\lambda =
\end{equation}
\[
\ \ \ \ \ \ \ \ \ \ \ \ =\log(K^*_V)-\log(M^*/L).
\]
 
The left side of this equation can be computed for each galaxy and
represents, for the proper values of $\log(\sigma)$ and $\muem$, the
difference between the observed FP and a fixed, reference VP [that for
which $\log(K^*_V)-\log(M^*/L)=0$]. In Section~\ref{sec4} we use such
a difference:

\begin{equation}\label{eqvirobs1}
\Delta_{FP}=(a-2)\log(\sigma)+(b-0.4)\muem+c-C_\lambda
\end{equation}
 
\noindent
to parametrize the position of galaxies onto the FP and we examine how
$\Delta_{FP}$ correlates with the observed proxies of the physical
factors to which the FP {\it tilt} could be ascribed.

\section{The WINGS dataset}\label{sec2}

The present work is based on two data samples extracted from the $WINGS$
database. The first one ({\it Sample I}) cross-matches the sample of
1550 ETGs used in Paper-I to study the FP in the V-band with the
galaxies in a subsample of 26 $WINGS$ clusters for which we obtained
$K$-band surface photometry \citep{Valentinuzzi,Donofrio}. In total we
got 620 ETGs. The second dataset ({\it Sample II}) contains 214 ETGs
and is obtained cross-matching the {\it Sample I} with the catalogues
of galaxy masses provided by \citet{Fritz}.

Likewise the $V$-band, the $K$-band surface photometry was obtained
using the purposely devised automatic software GASPHOT
\citep{Pignatelli}, which measured the total luminosity, the effective
radii, the mean effective surface brightness and the Sersic index of
each galaxy in the $WINGS$ clusters. These quantities were derived from
a simultaneous best-fit of the major and minor axes growth curves with
a Sersic law (\rnn\ ; \citealt{Sersic}) convolved with the local
$PSF$. The quality of the GASPHOT surface photometry is discussed in
\cite{Donofrio}, but several tests of its robustness can be found in
the papers already published by the $WINGS$ team (see \eg\
\citealt{Valentinuzzi1,Varela,Vulcani}).  The average uncertainties in the
surface photometry parameters are $\sim 10\%$, $\sim 20\%$ and $\sim
20\%$ for the luminosity, the effective radius and the Sersic index,
respectively. It is worth mentioning that the previously defined samples
do not include the galaxies for which the uncertainties of the best-fit
parameters exceeded three times the upper quartiles of the
corresponding distributions ($\sim 25\%$, $\sim 35\%$ and $\sim
35\%$ for the luminosity, the effective radius and the Sersic index,
respectively).

The procedure used to determine the stellar masses of galaxies from
the $WINGS$ spectra database \citep{Cava09} has been exhaustively
discussed by \cite{Fritz1,Fritz} and \cite{Vulcani}. Here we recall
that it is based on a spectrophotometric model that reproduces the
main features of observed spectra by summing the theoretical spectra
of simple stellar populations of different ages. Besides the stellar
masses, the tool is able to derive star formation histories, average
age and dust attenuation of galaxies. The models rely onto the Padova
evolutionary tracks \citep{Bertelli} and use the standard Salpeter
\citep{Salpeter} IMF, with masses in the range 0.15-120
$M_{\odot}$. Besides the stars which still are in the nuclear-burning
phase, the stellar mass includes remnants, such as white dwarves,
neutron stars and stellar black holes (for details see
\citealt{Fritz1}).  The stellar mass values relative to the whole
galaxy bodies are computed by rescaling to the total V-band magnitudes
the masses obtained by fitting the optical spectra, which are
calibrated on the V-band fiber magnitudes. Since this procedure
implicitly assumes there are no color gradients, we correct the
stellar masses following the prescriptions of \citet{BDJ} and using
the $\Delta(B-V)$ colors measured within the fiber aperture and at a
fixed aperture of 5 kpc. In \citet{Fritz} it is shown that the (total)
stellar masses computed in this way are in fairly good agreement with
other independent estimates, leading to an estimated accuracy of $\sim
0.2$ dex.

The galaxies were classified as early-types (Ellipticals or S0s) using
the automatic tool MORPHOT \citep{Fasano1}, purposely built for the
$WINGS$ project. For details about MORPHOT and about the accuracy
achieved for our morphological classification we refer to
\cite{Fasano1}, where an average $r.m.s.$ of $\sim$1.7 is reported for
the difference between automatic and visual estimates of the {\it
  Revised Hubble Type} (\citealt{deVauc}).

\begin{figure}
\begin{center}
\includegraphics[width=90mm]{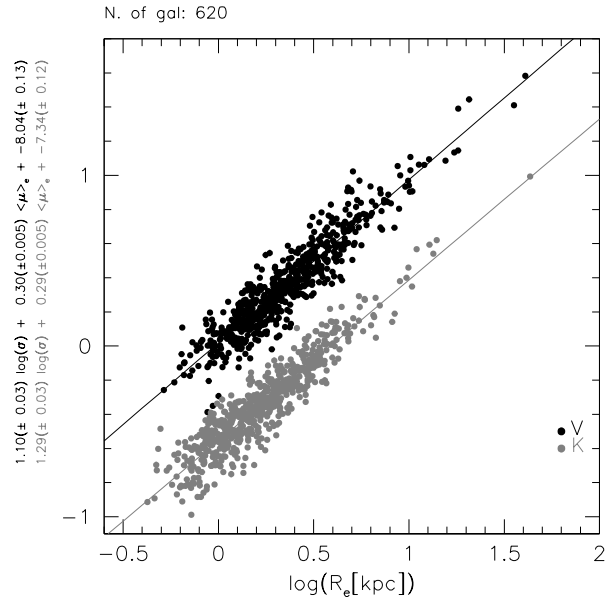}
\caption{
  The FPs for the ETGs of {\it Sample I} in the $V$- and $K$-bands are
  shown with black and grey dots respectively. The vertical shift of 0.5 is
  artificial. The derived coefficients are shown in the y-axis label.}
\label{Fig1_FP}
\end{center}
\end{figure}
  
The velocity dispersions have been taken from the literature
\citep{Smith,Bernardi}. They have been corrected for aperture effects
to $\re/8$ according to \citet{Jorg96} and have an average error of
$\sim 7-10$ \kms.

\section{The FP in the $V$ and $K$ bands}\label{sec3}

In Paper-I we studied the $V$-band FP of the ETGs members of the $WINGS$
clusters. Here we exploit the $K$-band extension of the $WINGS$ database
to derive a $K$-band solution for the FP coefficients. We remind that,
according to \citet{Donofrio1}, the FP coefficients do
depend on the wavelength of observations (see also
\citealt{Scodeggio,LaBarbera}), on the selection criteria of the
galaxy sample and on the environment \citep[see also
][]{Desroches,LaBarbera}.

In comparing the $WINGS$ $V$- and $K$-band FPs, we are allowed to ignore
the selection and environmental effects, since the same galaxies are
used in both wavebands. The fitting tool adopted in this work is the
same used in Paper-I, \ie\ the software MIST, kindly provided by
\citet{LaBarb}. We refer to their paper for any detail about the
fitting strategy adopted by MIST.  We obtain the following best-fit
FPs for our {\it Sample I} in the $V$- and $K$-bands:

\begin{equation}\label{ourFP}
\log(R^V_e)  = 1.10(\pm 0.03)\log(\sigma)+0.30(\pm 0.005)\langle\mu\rangle^V_e
\end{equation}
\[
-8.04(\pm 0.13),
\]

\begin{equation}\label{ourFPK}
\log(R^K_e) = 1.29(\pm 0.03)\log(\sigma)+0.29(\pm 0.005)\langle\mu\rangle^K_e \nonumber 
\end{equation}
\[
-7.34(\pm 0.12).
\]

In our formulation of the FP, \re\ is defined as the equivalent
(circularized) effective radius in $kpc$, $\sigma$ is the velocity
dispersion at $\re/8$ in \kms\ and $\langle\mu\rangle_e$ is the mean
effective surface brightness in mag\,arcsec$^{-2}$.  Note that, in
spite of the different samples used, the coefficients obtained here
for the $V$-band FP are quite close to the values found in Paper-I
($a=1.15\pm02$, $b=0.32\pm0.004$, $c=-8.56\pm0.09$; see that paper for
a discussion about the FP coefficients obtained using different data
samples). Note also that the FP slopes we find are a bit shallower
than those generally found in the literature. This systematic
difference, already discussed in Paper-I, is likely due to the fitting
procedure and to the magnitude limit of the data sample.

The FPs in the $V$- and $K$-band are shown in Figure~\ref{Fig1_FP} (see
eq.~\ref{ourFP} and \ref{ourFPK}). The scatter around the FP looks
very similar in the two bands ($r.m.s.\sim$ 0.092 and 0.088 for the
$V$- and $K$-band, respectively). Taking into account the individual
uncertainties of the three observed quantities (\re\, \muem\ and
$\sigma$), together with the covariance term involving the \re\ and
\muem\ uncertainties, we estimated the intrinsic scatter around the
$V$- and $K$-band FP to be 0.078 and 0.062, respectively, \ie\ nearly
half of the total observed scatter.

>From equations ~(\ref{ourFP}) and (\ref{ourFPK}) the FP {\it tilt}
increases going from the $K$- to the $V$-band. In particular, the gap
between the $\log\sigma$ coefficients in the $V$- and $K$-band turns
out to be $\sim0.2$, larger than the estimated fitting uncertainty and
in agreement with the value found by \cite{Pahre}.

It is worth noticing that, if we assume $\sigma$ as independent
variable, we obtain the following best-fit in the two bands:

\begin{equation}\label{ourFP1}
\log(\sigma)  = 0.72(\pm 0.02)\log(R^V_e)-0.21(\pm 0.006)\langle\mu\rangle^V_e
\end{equation}
\[
+6.18(\pm 0.12),
\]

\begin{equation}\label{ourFP1K}
\log(\sigma)  = 0.65(\pm 0.02)\log(R^K_e)-0.19(\pm 0.004)\langle\mu\rangle^K_e \nonumber 
\end{equation}
\[
+5.17(\pm 0.07).
\]

Again, the gap between the coefficients (relative to $\log R_e$ in
this case) in the $V$- and $K$-band turns out to be larger than the
estimated fitting uncertainty, thus implying that the difference has a
physical meaning.

\section{Origin of the FP tilt}\label{sec4}

\begin{figure*}
\centering
\vspace{-1truecm}
\includegraphics[width=140mm, angle=-90]{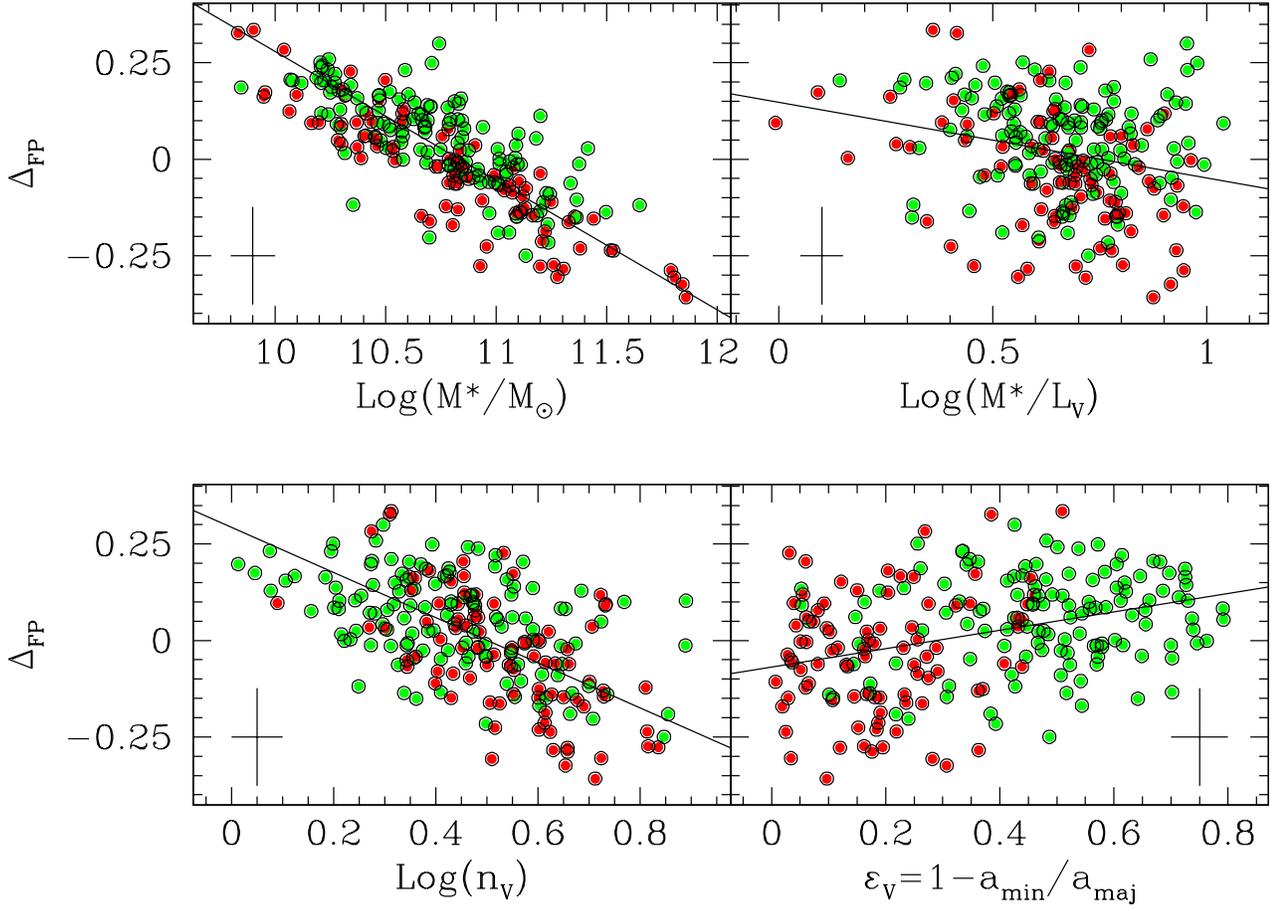}
\vspace{-0.5truecm}
\caption{$\Delta_{FP}$ in the $V$-band as a function of $\log(M^*/M_{\odot})$,
  $\log(M^*/L)$, $\log(n)$ and $\varepsilon$ for ETGs of
  $WINGS$-{\it Sample II}. Ellipticals
  and S0 galaxies are indicated by red and green dots,
  respectively. The straight lines illustrate the best-fit regressions
  (see text for details), while the crosses located aside in the
  plots report the median uncertainties of the two variables}
\label{Fig2_FP}
\end{figure*}
 
>From now on, unless warning of the contrary, the analysis of the $WINGS$
data will be based on {\it Sample II}.

Following the logical thread outlined in Section~\ref{sec1}, we adopt
the difference $\Delta_{FP}$ [see eq.~(\ref{eqvirobs1})] between the
FP and the reference VP defined therein, to parametrize the position
of the galaxy in the FP. Then, we analyze the systematic variation
along the FP of the different physical factors that might originate
the FP {\it tilt}.

\subsection{Structural non-homology and stellar population effects}\label{sec41}

\begin{table*}
\caption{Best-fit regression and Correlations coefficients of the relations 
$\Delta_{FP}$=$\alpha$X$+const.$ for the $V$-band $WINGS$ {\it Sample II}.}
\begin{tabular}{ccccccl}\\
\hline\hline
\multicolumn{7}{c}{V BAND DATA ($WINGS$ {\it Sample II})~~~~$\Delta_{FP}$=$\alpha$~$X$+$const.$   } \\
\hline
 $X$      &  $\alpha$  &   r.m.s.  & Pearson CC & Spearman CC & Significance & Sample \\
\hline
$\log(M^*/M_{\odot})$    &   -0.335  &   0.088   &  -0.800   &   -0.800   &  e-58  & ALL \\
$\log(M^*/L)$              &   -0.196  &   0.138   &   -0.225   &   -0.241   &  e-03 & ALL \\
$\log(n)$                     &  -0.585  &   0.122   & -0.541   &   -0.530   &   e-18  & ALL \\
$\varepsilon$              &   0.239  &    0.134   &   0.330   &    0.325   &   e-06  & ALL \\
\hline
$\log(M^*/M_{\odot})$    &   -0.325  &   0.072   &  -0.886   &   -0.880   &  e-38   & Es \\
$\log(M^*/L)$             &    -0.365  &   0.141  &  -0.400   &    -0.409  &   e-04  & Es \\
$\log(n)$                    &  -0.946  &   0.134   & -0.579   &   -0.563  &    e-09  & Es \\
$\varepsilon$             &    0.477  &    0.153  &   0.214   &    0.140   &   0.100  & Es \\
\hline
$\log(M^*/M_{\odot})$    &  -0.307  &   0.085   &  -0.731   &   -0.758   &  e-27   & S0s \\
$\log(M^*/L)$            &    -0.074  &   0.119  &   -0.105   &    -0.136  &   0.180 & S0s \\
$\log(n)$                   &  -0.375  &    0.108  &  -0.454  &   -0.416  &   e-07  & S0s \\
$\varepsilon$            &   0.113   &    0.118  &   0.152   &     0.107  &   0.160 & S0s \\
\hline\hline
\end{tabular}
\label{Tab0}
\end{table*}

In this subsection, by exploiting our homogeneous sample of ETGs, we
examine how $\Delta_{FP}$ correlates with $\log(M^*/L)$ (a proxy of
the stellar population) and with Sersic index and ellipticity ($n$ and
$\varepsilon=1-a_{min}/a_{maj}$). While the Sersic index is commonly
considered a robust proxy of the structural (non)homology, the role
of the ellipticity in this sense is more controversial.  We think
that, at least in a statistical sense, the ellipticity should be
considered, to all intents and purposes, a crucial ingredient of the
galaxy structure, thus being also deeply linked to the structural
(non)homology issue. In fact, is well known that different
(intrinsic) flattenings usually correspond in ETGs to different shapes
(oblateness vs. triaxiality) and luminosity profiles, even leaving
aside the obvious prevalence of the composite (bulge+disk) profiles at
increasing the ellipticity, due to the increasing fraction of S0
galaxies. In addition, since more flattened ETGs are preferentially
fast rotators \citep{binney,capp07,mccarthy,Emsell}, again in a
statistical sense, $\varepsilon$ should be linked to the amount of
rotational contribution to the total kinetic energy
($V_{rot}/\sigma$). On the other hand, it is hard to deny that a
different amount of rotation, by itself, is likely indicative of a
different dynamical structure.  Therefore, roughly speaking, we might
also consider $\varepsilon$ as a sort of indirect proxy of dynamical
(non)homology.

Figure~\ref{Fig2_FP} illustrates how $\Delta_{FP}$, computed from
eq.~(\ref{eqvirobs1}) in the $V$-band, correlates with stellar mass
and mass-to-light ratio ($M^*/L$), Sersic index ($n$) and ellipticity
($\varepsilon$). Ellipticals and S0 galaxies are represented in the
figure by red and green dots, respectively. The linear best-fits
reported in the plots have been computed through standard least square
regression analysis taking into account the individual uncertainties
on both variables (the crosses located aside in the plots outline the
median uncertainties). Table~\ref{Tab0} lists the parameters of both
the linear fits and the correlations in Figure~\ref{Fig2_FP} for the
whole {\it Sample II} and for Es and S0 galaxies separately.

Apart from the expected, strong correlation with the stellar mass, the
plots in this figure (see also Table~\ref{Tab0}) show that, among the
three quantities we tested for the FP {\it tilt}, the Sersic index is
the most strongly correlated with $\Delta_{FP}$ (i.e. with the
position along the FP), the correlations involving the ellipticity and
the stellar mass to light ratio (stellar populations) turning out to
be significantly weaker. This holds in both the $V$- and
the $K$-band (not reported here), at least for the ETGs sample as a
whole.
 
Besides the slopes and the correlation coefficients (both Pearson's
and Spearman's $CC$), what enforces the above conclusion are the
significances associated to the $CCs$ (see Table~\ref{Tab0}). These
are computed according to \citet{press92} and express the probability
that the null hypothesis of zero correlation is happening (small
values indicate significant correlations). Looking at the $CCs$ and
significances reported in Table~\ref{Tab0}, we note that, for the Es
alone, the $\Delta_{FP}$--$M^*/L$ correlation is much more significant
than for the S0 galaxies alone. We also note in Table~\ref{Tab0} that,
likely because of the reduced base-line, the correlation between
$\Delta_{FP}$ and $\varepsilon$ looks rather weak for Es and S0s
separately.

\begin{figure}
\centering
\vspace{-0.5truecm}
\includegraphics[width=90mm]{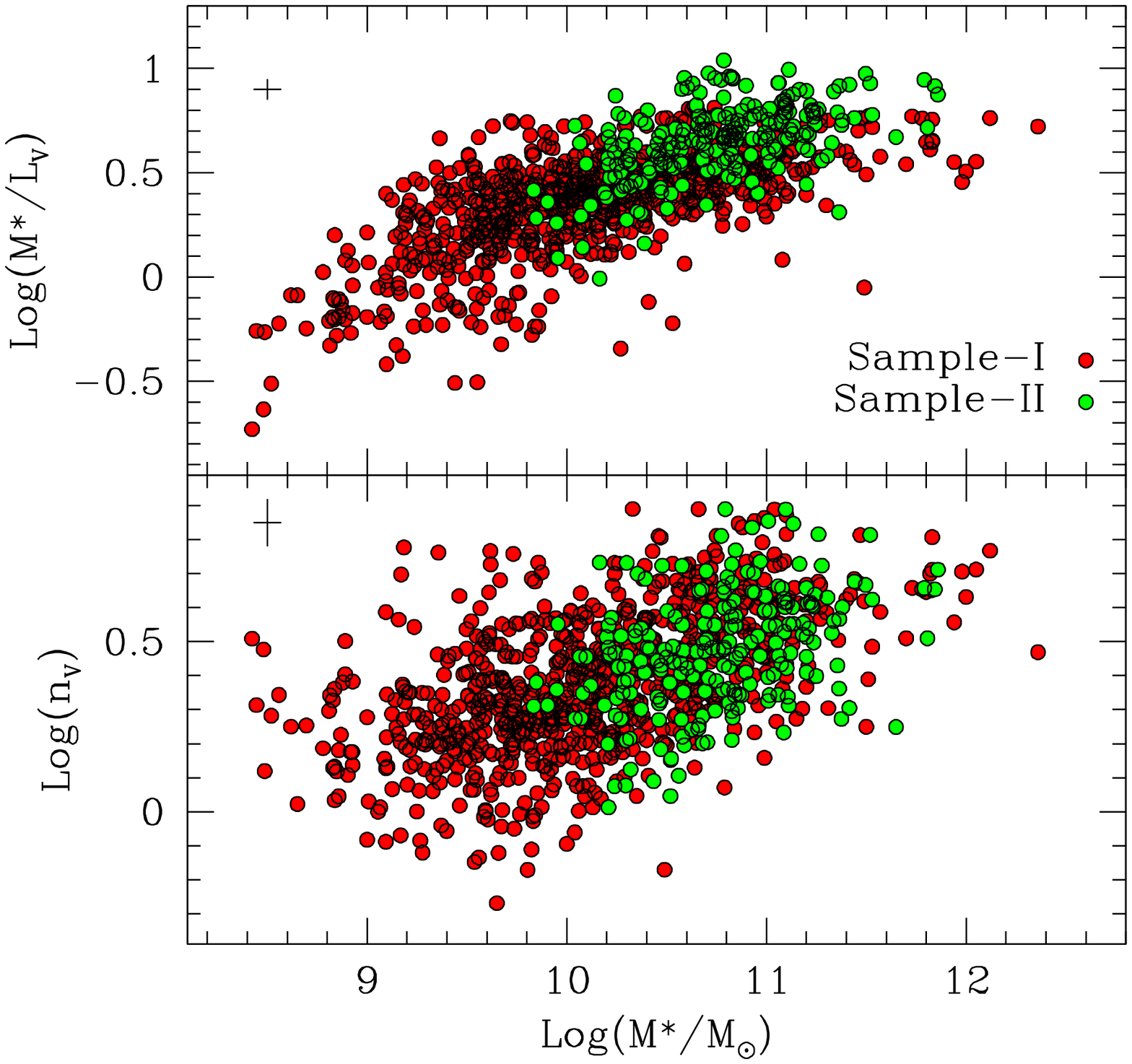}
\vspace{-0.5truecm}
\caption{The correlations of the galaxy stellar mass with $M^*/L_V$ (upper panel) and 
the Sersic index $n$ (lower panel) for the galaxies in our $Sample-I$ (red dots)
and $Sample-II$ (green dots).}
\label{Fig3_FP}
\end{figure}

Since many studies have found stellar population properties (e.g. age
and metallicity) to increase with galaxy stellar mass, it could look
surprising that $\Delta_{FP}$ turns out to be strongly correlated with
$M^*$, while the correlation between $\Delta_{FP}$ and $M^*/L$ is
weak. Does this mean that in our galaxy sample $M^*$ and
$M^*/L$ do not correlate? Furthermore, is there any internal
correlation among the four tested parameters ($M^*$, $M^*/L_V$, $n_V$
and $\varepsilon_V$)? Which is the relevant correlation in
Figure~\ref{Fig2_FP} ?

\begin{figure*}
\centering$
\begin{array}{ccc}
\includegraphics[width=57mm]{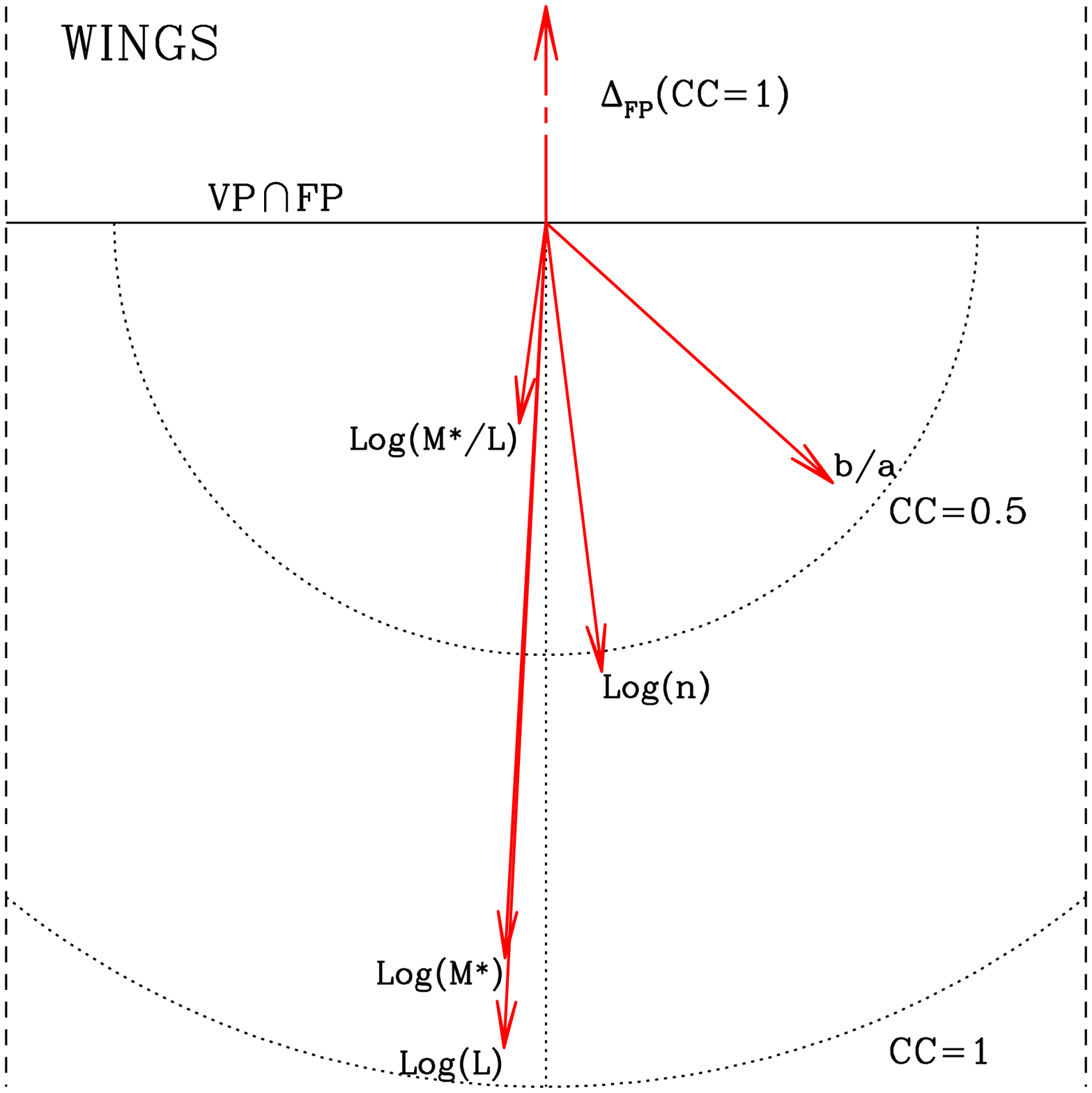} &
\includegraphics[width=57mm]{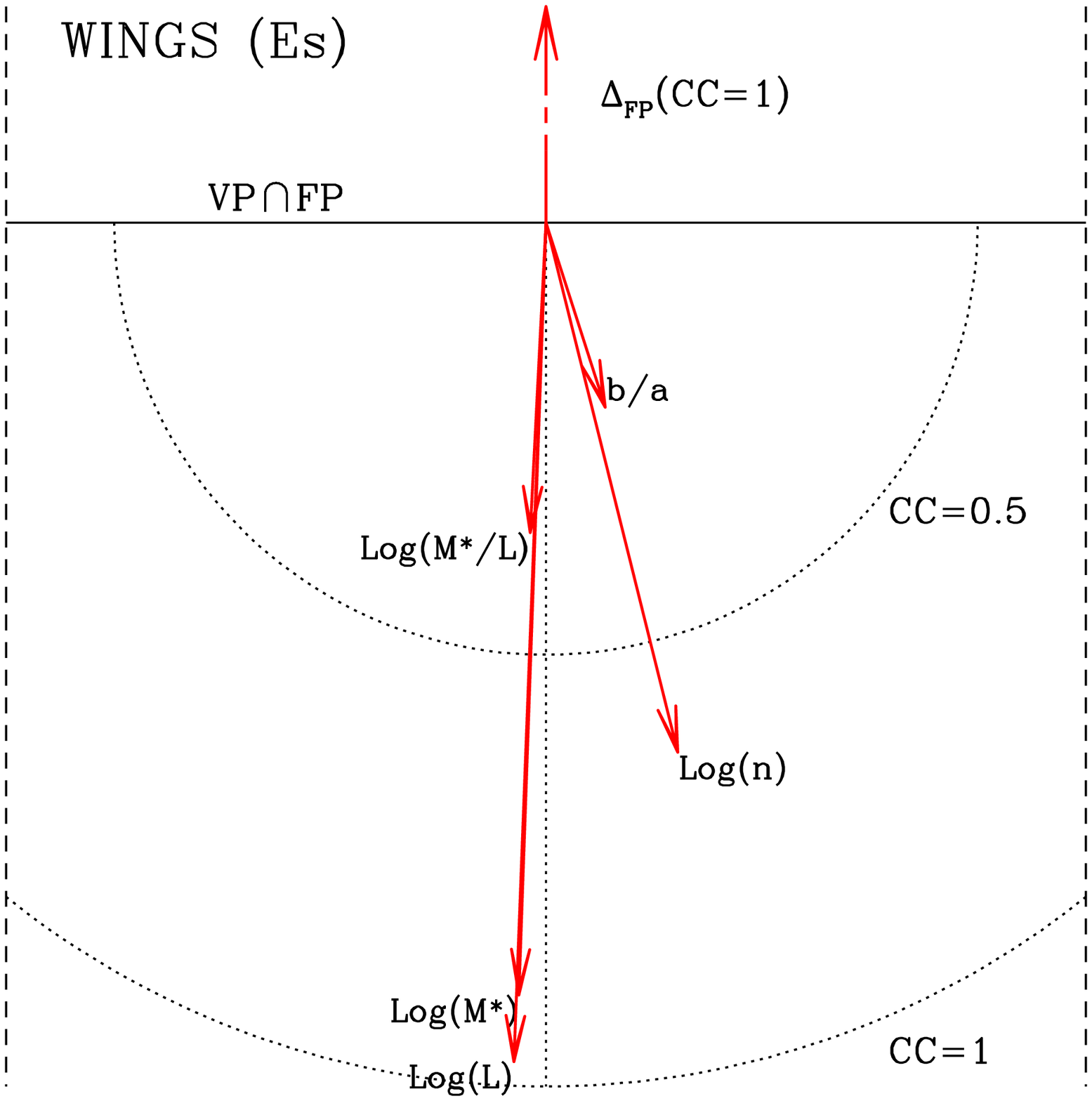} &
\includegraphics[width=57mm]{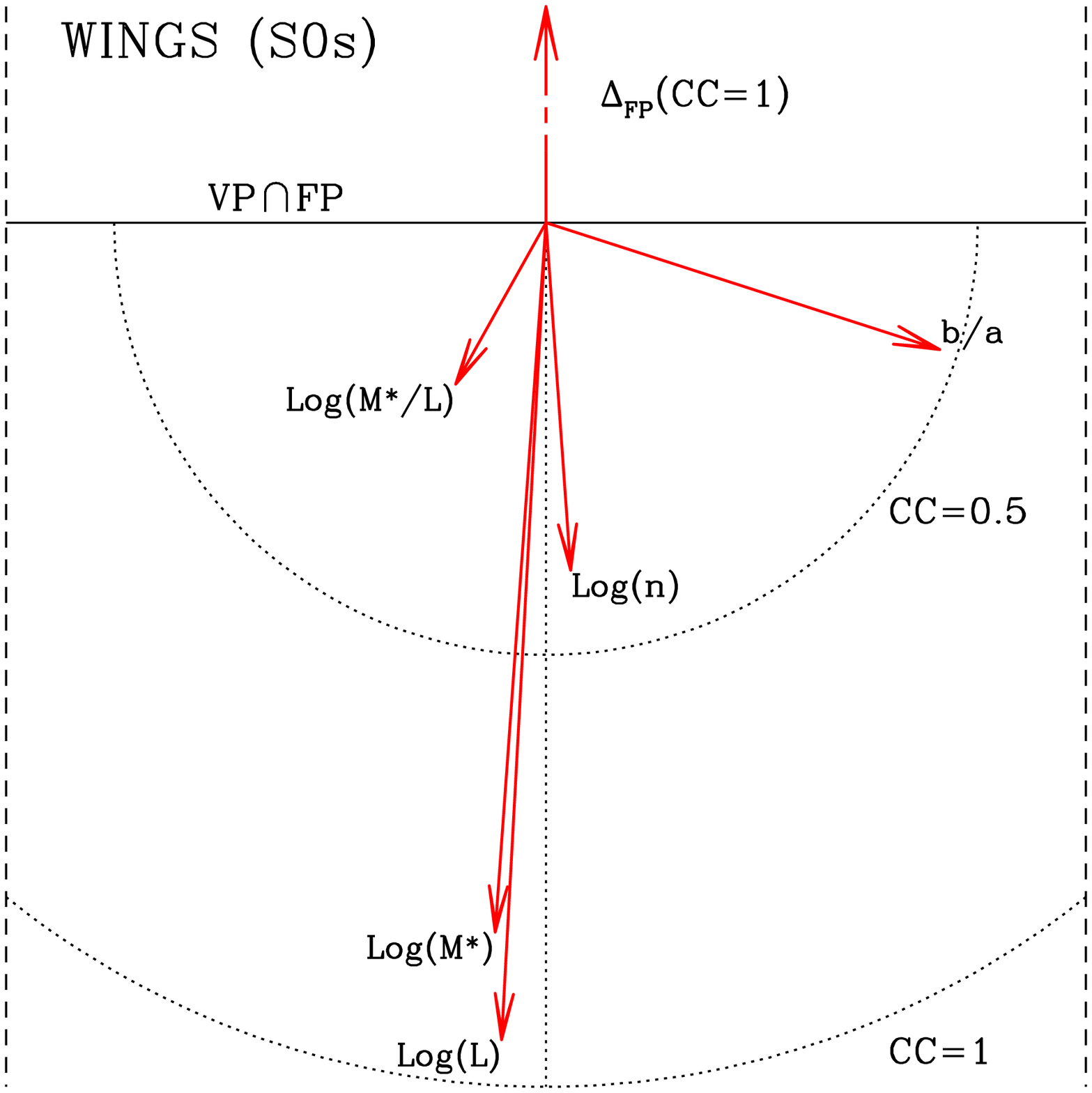}
\end{array}$
\caption{
  Face-on views of the V-band FP for our {\it Sample II} (left panel)
  and for the subsamples of ellipticals (central panel) and S0
  galaxies (right panel). The horizontal straight lines (VP$\cap$FP)
  represent the intersection between the FP and an arbitrary,
  reference VP. The red arrows in the figure indicate the directions
  of maximum correlation and the corresponding correlation strength
  for the various tested quantities ($M^*$, $L_V$, $M^*/L$, $n$ and
  b/a=1-$\varepsilon$). The dotted curves correspond to $CC=1$ and
  $CC=0.5$. Details about how the red arrows have been obtained are
  reported in the text.}
\label{Fig4_FP}
\end{figure*}

Figure~\ref{Fig3_FP} shows that, for our galaxy sample too, the
correlation between $M^*$ and $M^*/L$ is actually in place, being even
slightly more significant than that between $M^*$ and $n_V$ (0.66
vs. 0.55 of $CC$). In Figure~\ref{Fig4_FP} it is shown that such a
correlation does not translate into a strong correlation between
$M^*/L$ and $\Delta_{FP}$. This figure shows face-on
  views of the V-band FP for our {\it Sample II} (left panel) and for
  the subsamples of ellipticals and S0 galaxies (central and right
  panels, respectively). In each plot, the horizontal straight line
  marked with VP$\cap$FP represents the intersection between the FP
  (coefficients: $a$ and $b$) and an arbitrary, reference VP
  (coefficients: 2 and 0.4). Thus, the components of the direction
  vector of such stright line turn out to be: $b$-0.4, 2-$a$,
  $a$+2$b$. The red arrows in the figure lie onto the FP and, for each
  tested quantity ($M^*$, $L_V$, $M^*/L$, $n$ and
  b/a=1-$\varepsilon$), identify the direction along which it turns to
  show the strongest correlation. More precisely, spanning the whole
  round corner (0 to 2$\pi$), we rotate the coordinate axis along the
  FP and, for each angle, we compute the correlation coefficients of
  the various quantities, recording the directions which maximize
  them. The length of each arrow is proportional to the corresponding
  maximum value of $CC$ and the dotted curves correspond to $CC=1$ and
  $CC=0.5$. In this schematic picture, the arrow relative to
  $\Delta_{FP}$ must be perpendicular to the straight line VP$\cap$FP
  and must have size one by definition.

In Figure~\ref{Fig4_FP} both the stellar mass and the luminosity
appear to be strongly correlated with the position along the FP, the
direction of maximum correlation being in both cases almost
coincident, although with opposite direction, with that relative to
$\Delta_{FP}$ (upwards arrow with $CC$=1 by definition). Also the
Sersic index ($n$) appears quite well correlated with the position
along the FP (especially for Es) and substancially aligned with
$M^*$. The axial ratio (b/a) is generally correlated as well with the
position along the FP, but its behaviour depends on the morphological
type. In fact, while for Es there is a fair alignment with $M^*$ and a
weak correlation, for S0s there is a good correlation, but along a
direction almost orthogonal to $M^*$ (this is the reason why the $CCs$
in Table~\ref{Tab0} are small in both cases). In general, the stellar
mass-to-light-ratio ($M^*/L$) turns out to be less strongly correlated
with the position along the FP and, again, it shows a morphology
dependent behaviour. In this case, however, the correlation and the
alignment are coupled, both being stronger for Es than for S0s.

The Figure~\ref{Fig4_FP} should also help to clarify the question of
which are the truly important physical factors originating the FP {\it
  tilt}. The stellar mass is clearly the main driver of the FP {\it
  tilt} and drags the luminosity together. However, this is in some
sense an obvious thing, since we would like actually identify the '{\it
  mass-driven}' physical quantities which mainly contribute to shape
the FP through their influence on the observed parameters ($R_e$,
\muem ~and $\sigma$). To this concern, we checked that the
$\Delta_{FP}$ residuals relative to the $M^*-\Delta_{FP}$ relation do
correlate with both the Sersic index ($CC=0.31$) and the ellipticity
($CC=0.30$), the significance being very high in both cases ($\sim
10^{-6}$). 

\begin{figure}
\begin{center}
\includegraphics[width=90mm]{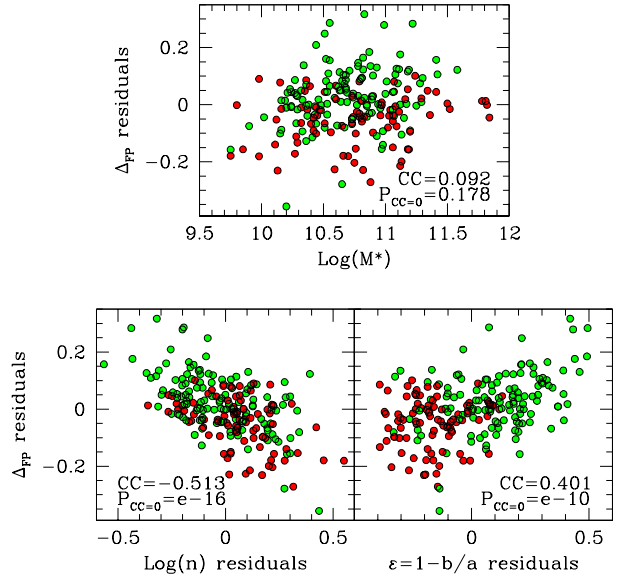}
\caption{
{\bf Upper panel}: residuals of the correlation $M^*-\Delta_{FP}$ (see upper-left panel of Fig.~\ref{Fig2_FP})
as a funtion of $M^*$.  {\bf Lower panels}: $M^*-\Delta_{FP}$ residuals as a function of the residuals of the $M^*-Log(n)$ and $M^*-\varepsilon$ correlations.}
\label{Fig5_FP}
\end{center}
\end{figure}

However, since all physical quantities strongly depend on the stellar
mass, the above mentioned correlations could be due to the possible
(spurious) correlation between $M^*-\Delta_{FP}$ residuals and $M^*$
itself. Thus, in order to properly check the existence of additional
(mass-free) dependences of $\Delta_{FP}$, we must look at the
correlations between $M^*-\Delta_{FP}$ residuals and the residuals of
the $M^*-Log(n)$ and $M^*-\varepsilon$ correlations. The lower panels
of Figure~\ref{Fig5_FP} illustrate these correlations, while in the
upper panel of the same figure the correlation between
$M^*-\Delta_{FP}$ residuals and $M^*$ it is shown. From the
correlation coefficients and significances reported in the figure, it
is clear that $M^*-\Delta_{FP}$ residuals do not correlate with $M^*$,
while the correlations with both the Sersic index and the ellipticity
residuals are even stronger than previously found for the
corresponding, mass-dependent quantities.  This led us to conclude
that both the Sersic index and the ellipticity are physical factors
actively driving the FP {\it tilt}.

\begin{table}
\caption{Multi-variate regression analysis coefficients of the
relation:
$\Delta_{FP}$=$\sum_{i=1}^3{c_i X_i}$+$const.$ for the $WINGS$ {\it Sample II}.}
\begin{tabular}{ccccc}\\
\hline\hline
    $X_i$             &   $c_i$  & $r.m.s.$ &  $t$ value & P($>\vert t\vert$) \\
\hline
$\log(M^*/L_V)$ & -0.115  &   0.041   &   -2.793   &   0.0057   \\
$\log(n)$           & -0.430  &   0.052   &   -8.294   &   e-14      \\
$b/a$                & -0.108  &   0.042   &   -2.540   &   0.0118   \\
\hline\hline
\end{tabular}
\label{MVRA}
\end{table}

Further support to this conclusion comes from both the Multi-Variate
Regression and the Principal Component Analyses (MVRA and PCA,
respectively).  Table~\ref{MVRA} reports the estimated MVRA
coefficients of the relation:
$\Delta_{FP}$=$c_1\log(M^*/L_V)$+$c_2\log(n)$+$c_3(b/a)$+$const.$,
together with the proper standard errors, the relative Student's $t$
statistics and the probabilities that the $t$ values are exceeded.
Table~\ref{MVRA} indicates that the coefficients of all three
variables are significant within the errors.  A similar indication
comes from the PCA, which is not able to reduce the number of
significant variables of the previous relation (including of course
$\Delta_{FP}$), all the four eigenvectors turning out to provide a
significant contribution to the total energy content of the relation
(49.4\%, 23.3\%, 17.0\% and 10.3\%, in descending order).

It is worth mentioning that all the results illustrated in this
Section remain unchanged when computing $\Delta_{FP}$ from the
coefficients of eq.~\ref{ourFP1} rather than those of eq.~\ref{ourFP},
\ie when using the velocity dispersion (instead of the effective
radius) as independent variable in the fit of the FP. The only
  noticeable difference with respect to the previous finding turns out
  to be the almost absent correlation between $\Delta_{FP}$ and
  $M^*/L$, with $CC=0.1, 0.25$~and~$-0.06$ [P($>\vert t\vert$)$\sim
  0.19, 0.02$ and $0.5$] for the global, elliptical and S0 samples,
  respectively. This might indicate that the role of stellar
  populations in determining the FP {\it tilt} is marginal, as the
  strength of the correlation between $M^*/L$ and $\Delta_{FP}$ seems
  to depend on the fitting procedure.

We believe that the above findings strongly enough support the
conclusion that various physical quantities concur in shaping the
FP. Still, in Section~\ref{sec7} we show evidences suggesting that the
FP properties ({\it tilt} and thickness) altogether could be
interpreted as due to the existence of a connection between stellar
mass, structure and stellar populations in ETGs \citep[nMML
relation;][]{Donofrio2}. In this framework, it would be misleading to
attribute the FP properties to the mere summing of the influences of
the individual physical quantities, since they are actually entangled
through the nMML relation.

\subsection{Comparison with ATLAS3D and SDSS data}\label{sec42}

In order to check the robustness of the results obtained in the
previous section from the analysis of the $WINGS$ data sample, we
perform here a similar analysis using the $ATLAS3D$ Project data
\citep[][C11 hereafter]{Cappellari2011} and a subsample of the
$SDSS$ and follow-up data \citep{Abaza09}. This will allow us to
compare with our results and to speculate about the influence of
dynamical non-homology and dark matter fraction on the FP {\it
  tilt}. Both the $ATLAS3D$ and the $SDSS$ magnitudes are corrected
for galactic extinction and are given in the $r$-band, AB system.

Among the 260 galaxies of the $ATLAS3D$ sample, we selected 232
galaxies classified as early-type (Es and S0s) by C11. For these
galaxies, we took from \citet{Krajn13} the Sersic index ($n$), derived
fitting the $r$-band luminosity profiles with a single Sersic law, and
from \citet{Cappellari4} the dynamical mass-to-light-ratio
($M_{Tot}/L$), the FP parameters (\re , $L/L_{\odot}$ and $\sigma_e$)
and the ellipticity ($\varepsilon_e$). All these quantities are
derived using the JAM dynamical model \citep{capp08} and the MGE
surface photometry tool \citep{Emsell94}. To improve the robustness of
the analysis, we decided to remove from the sample those galaxies
  for which the effective radius and the Sersic index uncertainties
  exceed 2.5$\times rms$ of the relative distributions, as well as the
  galaxies with extremely faint surface brightness or very small
  velocity dispersion.  We also remove from the sample those galaxies
  for which the quality flag $q$ of the JAM fitting given by
  \citet{Cappellari4} is zero (bad fitting).  In particular, we kept
in the final $ATLAS3D$ sample just those galaxies obeying the
following conditions: $n\le$8, $\log(\sigma_e)\ge$1.6, $\muem\le$24,
$\delta n\le$1.5, $\delta n/n\le$0.4, $\delta \re/\re\le$0.3 and
$q>$0, where $\delta n$ and $\delta \re$ are the uncertainties of $n$
and \re, respectively.

For the remaining sample of 137 $ATLAS3D$ galaxies, we
took from \citet{Cappellari3} the JAM model stellar
mass-to-light-ratio ($M^*/L$) and from \citet{Emsell} the ratio
$V_{rot}/\sigma_e$ between the rotation speed and the velocity
dispersion. The distances of the $ATLAS3D$ galaxies are taken from C11
and are always less than 45~Mpc.

The sample of 817 $SDSS$ galaxies we use here ($WSDSS$
  hereafter) is obtained cross-matching the whole $SDSS-DR7$ galaxy
sample with the galaxies classified as early-type (Es and S0s) in the
$WINGS$ database. The surface photometry parameters in the
  $r$(AB)-band, including the Sersic index $n$, $\varepsilon$, \re
~and \muem ~are taken from \citet{Blant03} and are derived fitting the
azimuthally averaged radial luminosity profiles of galaxies with a
single component Sersic law. The central velocity dispersions
($\sigma_c$) and the ellipticities ($\varepsilon$) come from the
$SDSS$ database \citep{Abaza09}. The stellar masses $M^*$ are taken
from the MPA-JHU DR7 release of spectrum measurements
($http://www.mpa-garching.mpg.de/SDSS/DR7/$) and are obtained by
fitting the population synthesis model of \citet{Bruz03} to the $SDSS$
broad bands $u,g,r,i,z$ magnitudes adopting the universal initial mass
function (IMF) as parametrized by \citet{Kroupa}.  The total
  luminosities in th V-band have been obtained from the $r$- and
  $g$-band total magnitudes from \citet{Blant03} using the recipe of
  \citet{Lupt}. After conversion to the Vega-mag system, they have
  been used to compute the mass-to-light ratios $M^*/L_V$.  Finally,
the distance moduli and the morphological types are taken from the
$WINGS$ database.

For both the $ATLAS3D$ and the $WSDSS$ samples, we have obtained the
best fitting of the FP in the $r$-band using the same fitting
algorithm used for the $WINGS$ data (MIST). We obtained the following
equations:

\begin{equation}\label{ATLASFP}
\log(R_e)  = 1.31(\pm 0.05)\log(\sigma)+0.33(\pm 0.009)\langle\mu\rangle_e
\end{equation}
\[
-8.58(\pm 0.22),
\]

\begin{equation}\label{SDSSFP}
\log(R_e) = 1.17(\pm 0.05)\log(\sigma)+0.32(\pm 0.009)\langle\mu\rangle_e
\end{equation}
\[
-8.12(\pm 0.37),
\]

\noindent
for the $ATLAS3D$ and the $WSDSS$ samples, respectively.

 While the FP coefficients relative to the $WINGS$ sample fairly
  agree with $WSDSS$, they turn out to be significantly
  different from those relative to the $ATLAS3D$ sample. Apart from
  the different wavebands (V for $WINGS$ and $r$ for $ATLAS3D$) and
  luminosity functions of the two galaxy samples, this difference
  might originate from the different methods used to measure the
  quantities involved in the FP (\re, \muem\ and $\sigma$). Since it is
  well known \citep[see for instance][]{Kelson} that the correlated
  variation of \re and \muem from different fitting methods has
  negligible effects on the FP coefficients, we guess the differences
  we found in this case are caused by differences in the methods used
  to measure velocity dispersions.  Actually, while the $\sigma$ of
  the $WINGS$ galaxies are normalized according to the canonical rule
  proposed by \citet{Jorg96} ($R_e/8$), those given in $ATLAS3D$ are
  averaged within the effective radius.

\begin{figure}
\begin{center}
\includegraphics[width=100mm]{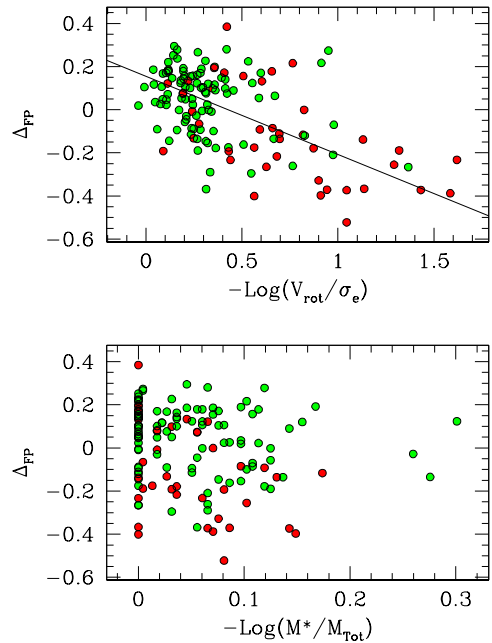}
\caption{
$\Delta_{FP}$ as a function of $\log(V_{rot}/\sigma_e)$ and $\log(M^*/M_{Tot})$
for 137 galaxies in the ATLAS3D sample. Symbols are as in the previous figures.}
\label{Fig6_FP}
\end{center}
\end{figure}

\begin{figure*}
\begin{center}$
\begin{array}{cc}
\includegraphics[width=90mm]{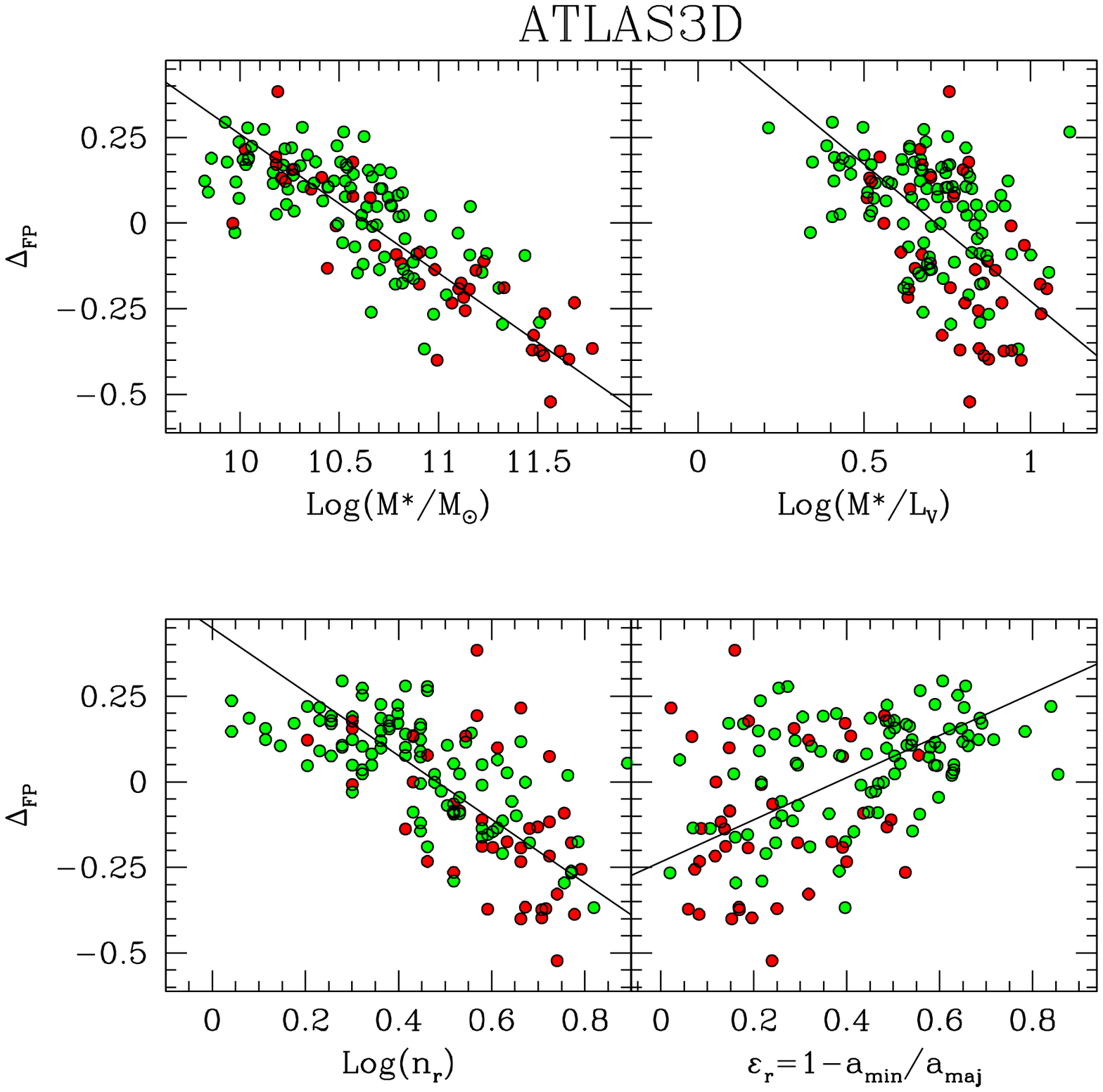} &
\includegraphics[width=90mm]{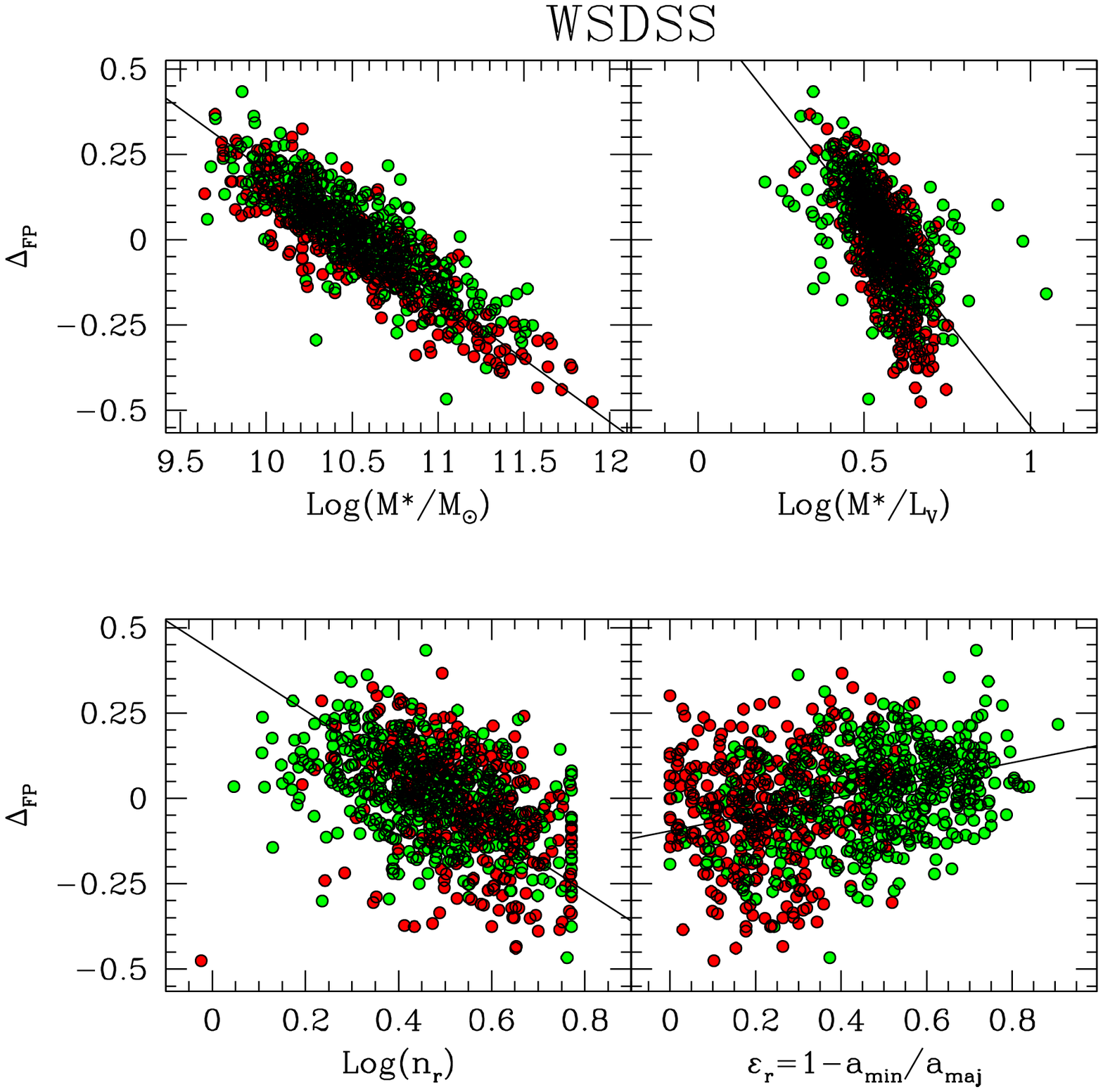}
\end{array}$
\vspace{-0.5truecm}
\caption{
  {\bf Left panels}: $\Delta_{FP}$ in the $V$-band as a function of
  $\log(M^*/M_{\odot})$, $\log(M^*/L)$, $\log(n)$ and $\varepsilon$
  for 137 ETGs of The ATLAS3D sample. Symbols are as in
  Fig~\ref{Fig2_FP}.  {\bf Right panels}: same plots for 817 ETGs in
  common between $SDSS$(DR7) and $WINGS$}
\label{Fig7_FP}
\end{center}
\end{figure*}

As in the case of the $WINGS$ data, through the eq.~(\ref{eqvirobs1}),
we computed the quantities $\Delta_{FP}$ for the galaxies of the two
samples and correlated them with the various physical parameters
supposed to be responsible of the FP {\it tilt}. The results are
illustrated in the Figures~\ref{Fig6_FP} and \ref{Fig7_FP} and in the
Tables~\ref{Tabatlas} and \ref{Tabsdss}. Note that, in order to
compare with our results about the stellar mass-to-light-ratio, we
converted the $SDSS$ stellar masses from Kroupa to Salpeter IMF adding
0.13 dex and reported in Fig~\ref{Fig7_FP} the $M/L$ in the
$V$-band. For the conversion from the $r$(AB)- to the $V$-band we used
the equations given in \citet{Blant07} and the $g$ and $r$ magnitudes
from $SDSS$. For the 41 $ATLAS3D$ galaxies not included in the $SDSS$
database, we used the (V-R) colors from NED.

\begin{table*}
\caption{Best-fit regression and correlations coefficients of the relations 
$\Delta_{FP}$=$\alpha$X$+const.$ from the ATLAS3D data.}
\begin{tabular}{ccccccl}\\
\hline\hline
\multicolumn{7}{c}{ATLAS3D data sample~~~~$\Delta_{FP}$=$\alpha$~$X$+$const.$   } \\
\hline
$X$      &  $\alpha$  &   r.m.s.  & Pearson CC & Spearman CC & Significance & Sample \\
\hline
$\log(M^*/M_{\odot})$    &   -0.406  &   0.114   &  -0.813   &   -0.804   &  e-39  & ALL \\
$\log(M^*/L)$              &   -0.799  &   0.171   &   -0.629   &   -0.622   &  e-09 & ALL \\
$\log(n)$                     &  -0.930  &   0.147   & -0.664   &   -0.689   &   e-22  & ALL \\
$\varepsilon$              &   0.616  &    0.171   &   0.444   &    0.433   &   e-08  & ALL \\
$-\log(V_{rot}/\sigma_e)$    &   -0.362  &   0.157   &  -0.546   &   -0.429   &  e-10  & ALL \\
$-\log(M^*/M_{Tot})$    &   0.008  &   0.184   &  0.170   &   0.178   &  0.040  & ALL \\
\hline
$\log(M^*/M_{\odot})$    &   -0.406  &   0.100   &  -0.891   &   -0.889   &  e-19  & Es \\
$\log(M^*/L)$             &    -1.390  &   0.209  &  -0.512   &    -0.535  &   e-04  & Es \\
$\log(n)$                    &  -1.660  &   0.218   & -0.565   &   -0.566  &    e-06  & Es \\
$\varepsilon$             &   -0.364  &    0.229  &   0.161   &    0.159   &   0.310  & Es \\
$-\log(V_{rot}/\sigma_e)$    &   -0.376  &   0.176   &  -0.585   &   -0.594   &  e-05  & Es \\
$-\log(M^*/M_{Tot})$    &   -0.358  &   0.216   &  -0.401   &   -0.417   &  0.006  & Es \\
\hline
$\log(M^*/M_{\odot})$    &  -0.386  &   0.116   &  -0.694   &   -0.716   &  e-18   & S0s \\
$\log(M^*/L)$            &    -0.418  &   0.138  &   -0.384   &    -0.399  &   e-04 & S0s \\
$\log(n)$                   &  -0.702  &    0.116  &  -0.655  &   -0.665  &   e-14  & S0s \\
$\varepsilon$            &   0.399   &    0.135  &   0.426   &     0.394  &   e-05 & S0s \\
$-\log(V_{rot}/\sigma_e)$    &  -0.271  &   0.145   &  -0.285   &   -0.237   &  0.011  & S0s \\
$-\log(M^*/M_{Tot})$    &   0.137  &   0.148   &   0.140   &   0.141   &  0.167  & S0s \\
\hline\hline
\end{tabular}
\label{Tabatlas}
\end{table*}

\begin{table*}
\caption{Best-fit regression and correlations coefficients of the relations 
$\Delta_{FP}$=$\alpha$X$+const.$ from $WSDSS$ data}
\begin{tabular}{ccccccl}\\
\hline\hline
\multicolumn{7}{c}{$WSDSS$(DR7) data ~~~~$\Delta_{FP}$=$\alpha$~$X$+$const.$   } \\
\hline
$X$      &  $\alpha$  &   r.m.s.  & Pearson CC & Spearman CC & Significance & Sample \\
\hline
$\log(M^*/M_{\odot})$    &   -0.367  &   0.080   &  -0.848   &   -0.833   &  0.000  & ALL \\
$\log(M^*/L)$              &   -1.230  &   0.118   &   -0.603   &   -0.664   &  e-90 & ALL \\
$\log(n)$                     &  -0.880  &   0.144   & -0.441   &   -0.468   &   e-45  & ALL \\
$\varepsilon$              &   0.249  &    0.141   &   0.286   &    0.287   &   e-16  & ALL \\
\hline
$\log(M^*/M_{\odot})$    &   -0.363  &   0.077   &  -0.888   &   -0.873   &  0.000   & Es \\
$\log(M^*/L)$             &    -1.950  &   0.124  &  -0.681   &    -0.658  &   e-47  & Es \\
$\log(n)$                    &  -1.450  &   0.190   & -0.409   &   -0.486  &    e-18  & Es \\
$\varepsilon$             &    0.561  &    0.173  &   0.104   &    0.095   &   0.077  & Es \\
\hline
$\log(M^*/M_{\odot})$    &  -0.359  &   0.076   &  -0.828   &   -0.834   &  0.000   & S0s \\
$\log(M^*/L)$            &    -0.914  &   0.107  &   -0.575   &    -0.668  &   e-60 & S0s \\
$\log(n)$                   &  -0.675  &    0.124  &  -0.432  &   -0.426  &   e-24  & S0s \\
$\varepsilon$            &   0.257   &    0.124  &   0.320   &     0.308  &   e-12 & S0s \\
\hline\hline
\end{tabular}
\label{Tabsdss}
\end{table*}

Figure~\ref{Fig7_FP} and Tables~\ref{Tabatlas} and \ref{Tabsdss} allow
a direct comparison with the results obtained for the $WINGS$ sample
(see Fig.~\ref{Fig2_FP} and Table~\ref{Tab0}). From this comparison, we
draw the following general conclusions: {\it (i)} for the stellar mass
and the structural parameters $n$ and $\varepsilon$, both relative to
the whole samples (Es+S0s) and to the Es and S0s samples separately,
the $ATLAS3D$ and $WSDSS$ data produce correlations very similar to the
corresponding ones we found for the $WINGS$ galaxy sample. Actually, for
both $n$ and $\varepsilon$, the $CCs$ from $WINGS$ are intermediate
between those obtained from the two comparison samples; {\it (ii)} the
correlation between $\Delta_{FP}$ and $M^*/L_V$ turns out to be much
stronger for the $ATLAS3D$ and $WSDSS$ samples than for the $WINGS$
sample, although for the $WINGS$ Es alone the correlation is quite
significant as well.

One could speculate that such discrepancy is because the origin of the
tilt and scatter of the FP might be not the same for galaxies residing
in different environments (i.e.  $WINGS$ vs.  $WSDSS$ and $ATLAS3D$
samples). However, even though the dependence of the FP
coefficients on the environment still is a controversial issue
\citep{Bernardi,LaBarbera}, many works have shown the FP tilt and
scatter to be nearly the same in different environments
\citep{Jorg96,Pahrea,Pahre,Kocha,delarosa,Magoulas}. Since it is
hard to believe that different physical mechanisms operating in
different environments produce the same FP shape, we are inclined to
believe that the origin of the above discrepancy lies in the
different recipes adopted to calculate the total stellar mass of the
galaxies. We do not enter here in the ample literature debate
concerning the theoretical models which better reproduce the
properties of the stellar population. We only note that: {\it a)} the
range of $M^*/L_V$ values found with the $WSDSS$ data is smaller than
those provided by $WINGS$ and $ATLAS3D$ (see Fig.~\ref{Fig7_FP}); {\it
  b)} the $ATLAS3D$ $M^*/L$ ratios are measured within 1\re, so they
cannot be directly compared with those of $WINGS$ and $WSDSS$, in
particular if one recognizes that strong color gradients exist in ETGs
of high mass (see below Fig.~\ref{Fig13_FP}).  Taking into account
that the $WINGS$ masses are obtained from the analysis of the whole
SED of our galaxies and are corrected for color gradients, at variance
with those derived by $ATLAS3D$ that are based on the line strength
index of $H_{\beta}$ \citep{Cappellari4}, we are confident that our
masses and $M^*/L_V$ measurements are quite robust; {\it c)} the
observed discrepancy will remain an open problem until new more
reliable stellar mass measurements will be available for galaxies.

>From Figure~\ref{Fig6_FP} and Table~\ref{Tabatlas} (just $ATLAS3D$
data), we conclude that: {\it (iii)} $M^*/M_{Tot}$ significantly
correlates with $\Delta_{FP}$ just for the Es sample; {\it (iv)} the
correlation between $\Delta_{FP}$ and $V_{rot}/\sigma_e$ turns out to
be very significant for Es and, to a lesser extent, for S0 galaxies.

\begin{figure}
\begin{center}
\includegraphics[width=90mm]{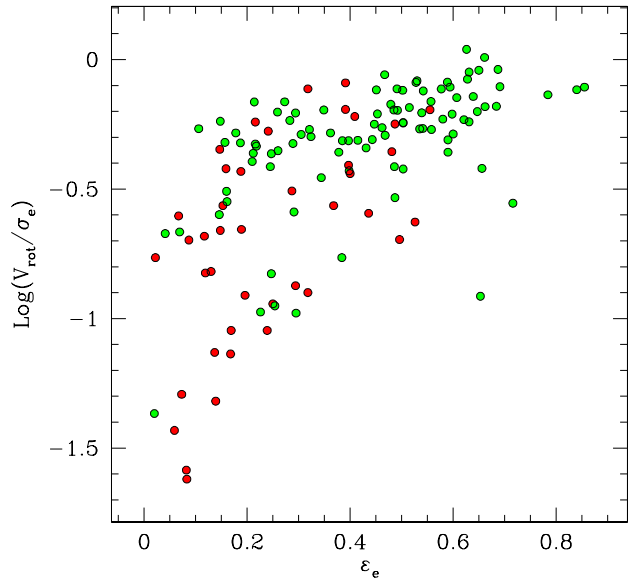}
\caption{
  The correlation between the ellipticity $\varepsilon_e$ and
  $V_{rot}/\sigma_e$ for the $ATLAS3D$ galaxy sample. 
  Symbols are as in the previous figures.}
\label{Fig8_FP}
\end{center}
\end{figure}

Concerning the last point, in the first paragraph of Sec.~\ref{sec41}
we briefly discussed the role played by both $\varepsilon$ and
$V_{rot}/\sigma$ in the structural and dynamical (non)homology
issue. We argued that, at least in a statistical sense, the
ellipticity is likely involved in both the luminous and dynamical
structure of ETGs. We also argued that, even if specifically
indicating the rotational contribution to the kinetic energy, the
quantity $V_{rot}/\sigma$ could also be thought as a rough diagnostic
(proxy) of the dynamical structure of ETGs. Here, the similarity of
the correlations linking $\Delta_{FP}$ with $\varepsilon$ and
$V_{rot}/\sigma$ for the $ATLAS3D$ sample (see left panels of
Figure~\ref{Fig7_FP} and upper panel of Figure~\ref{Fig6_FP}), suggests
to conclude that, in some sense, $\varepsilon$ and $V_{rot}/\sigma$
could be though as tokens of the same physical phenomenon. Indeed, in
Figure~\ref{Fig8_FP} it is shown that a strong correlation exists between
$\varepsilon$ and $V_{rot}/\sigma$. A question could be raised: which
of the two quantities is driving the correlation with $\Delta_{FP}$?
We tried to answer this question through the analysis of the
residuals, but our attempts were unsuccessful, since the
CCs and the significances of the residual correlations turned out to
be always inconclusive, although for the Es the residuals of the
$V_{rot}/\sigma$-$\Delta_{FP}$ regression turn out to be marginally
correlated with $\varepsilon$ (significance $\sim 0.07$). In the last
part of this sub-section it is shown that the combined MVRA and PCA 
results seem to favour $V_{rot}/\sigma$ as driver quantity of the
correlation with $\Delta_{FP}$.

\begin{figure*}
\begin{center}$
\begin{array}{ccc}
\includegraphics[width=57mm]{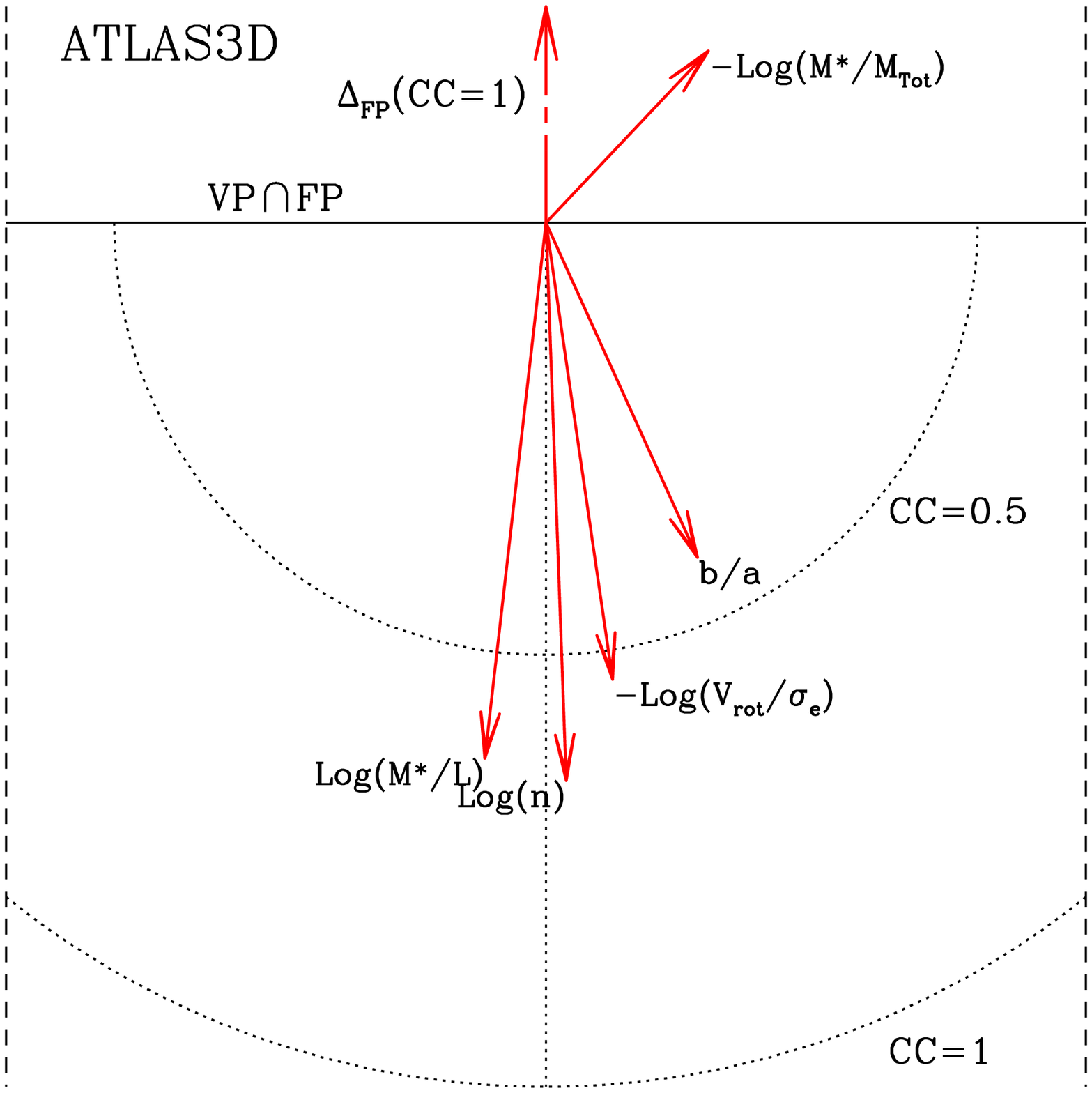} &
\includegraphics[width=57mm]{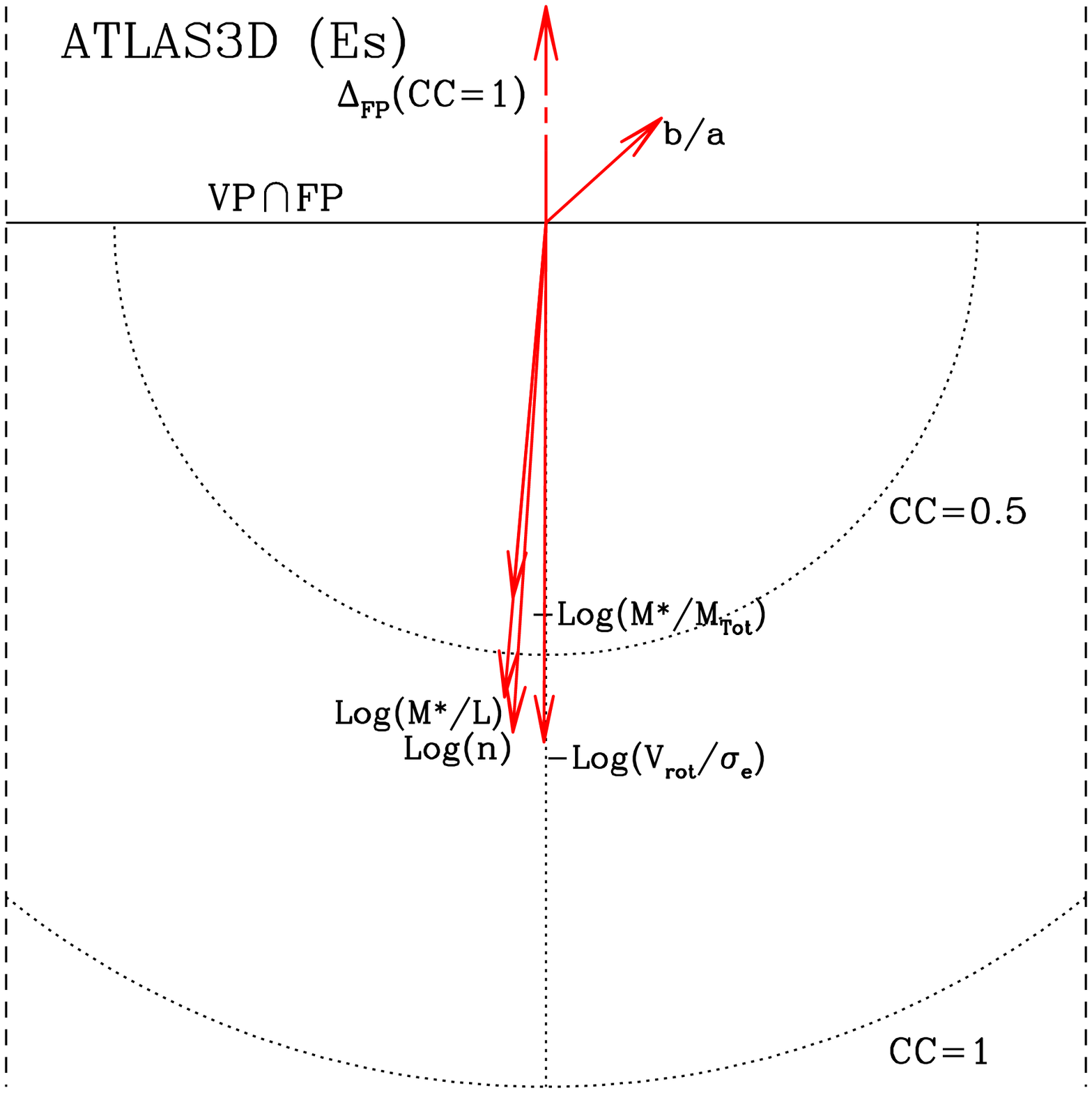} &
\includegraphics[width=57mm]{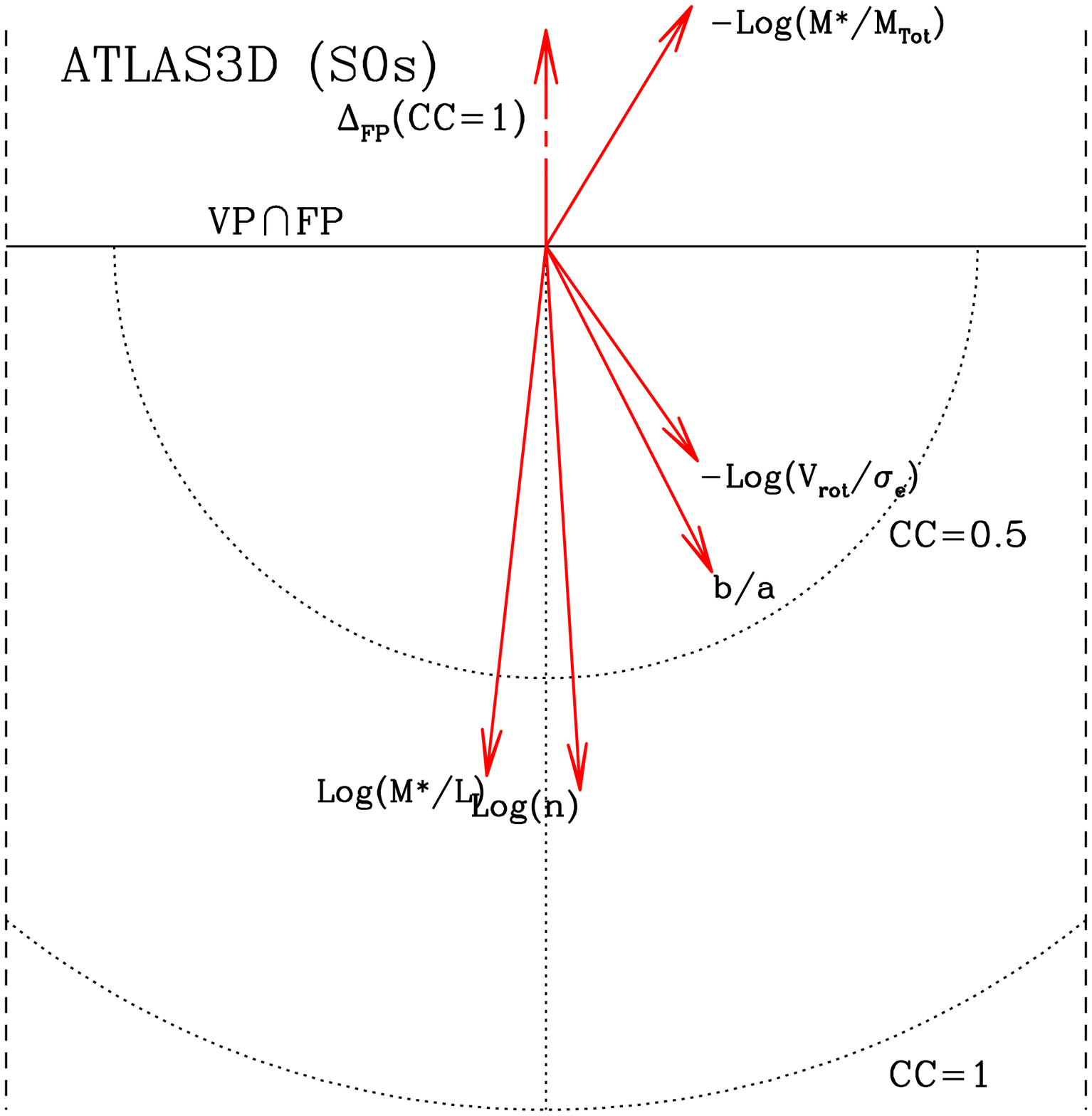}
\end{array}$
$\begin{array}{ccc}
\includegraphics[width=57mm]{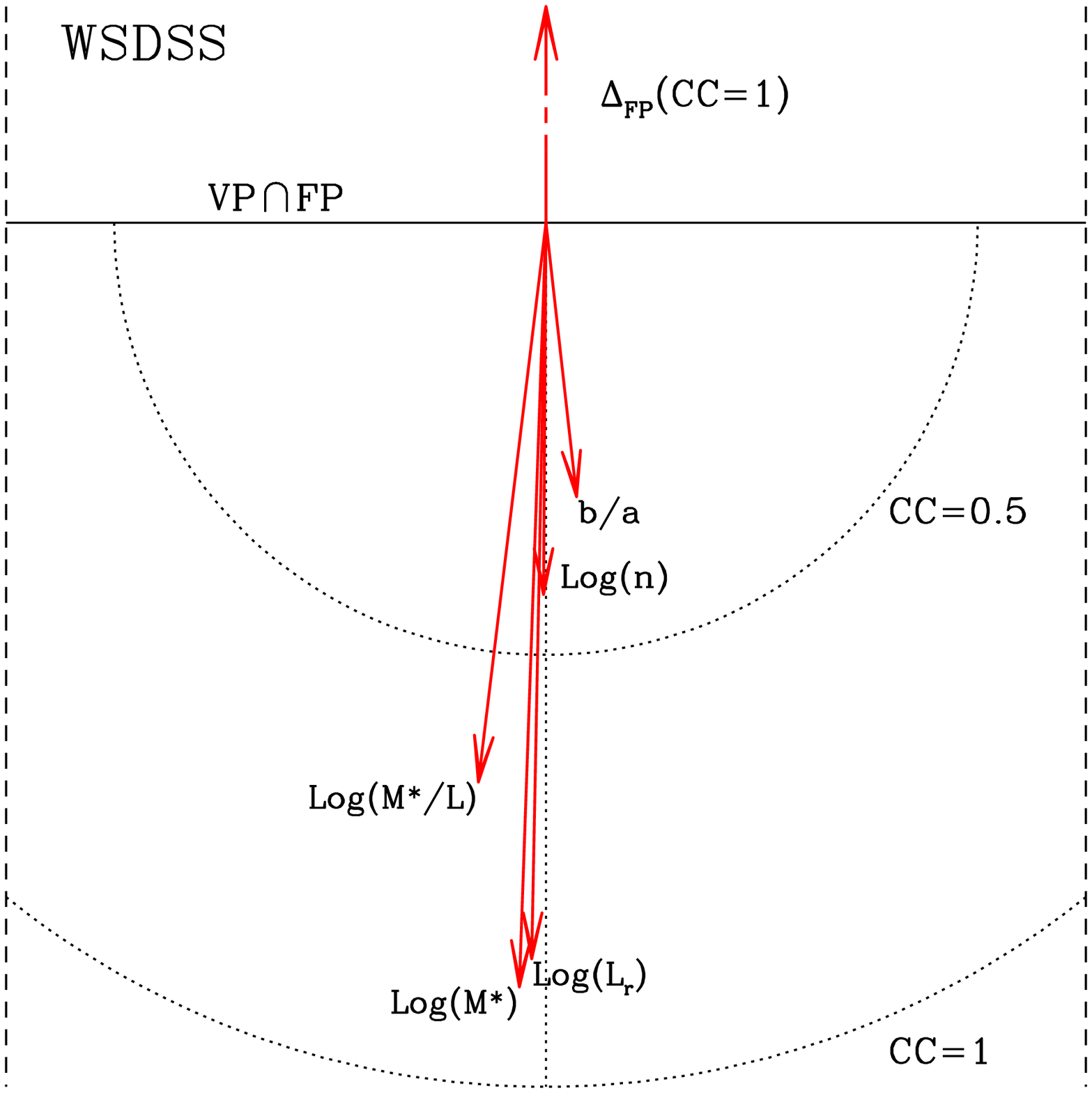} &
\includegraphics[width=57mm]{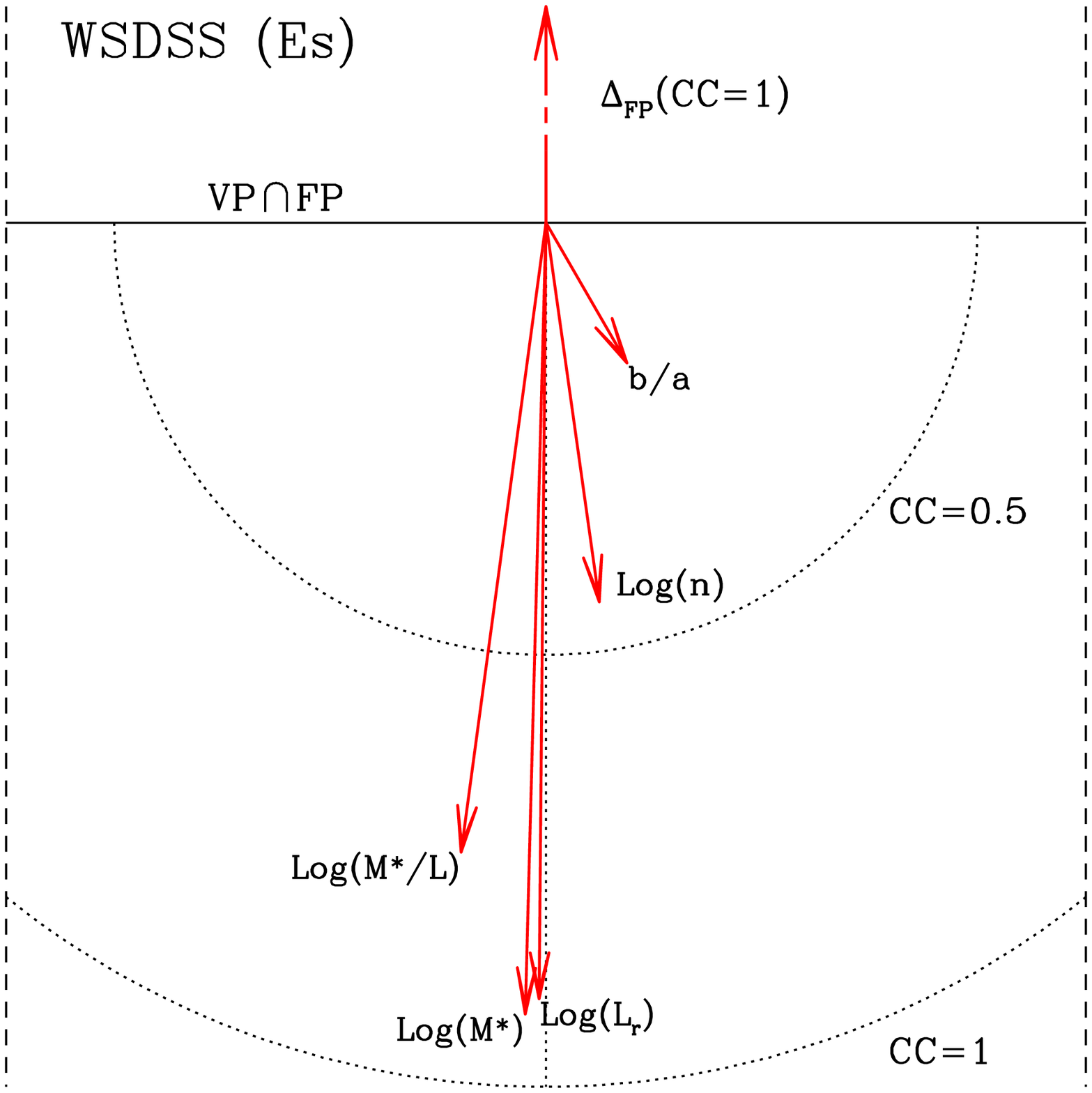} &
\includegraphics[width=57mm]{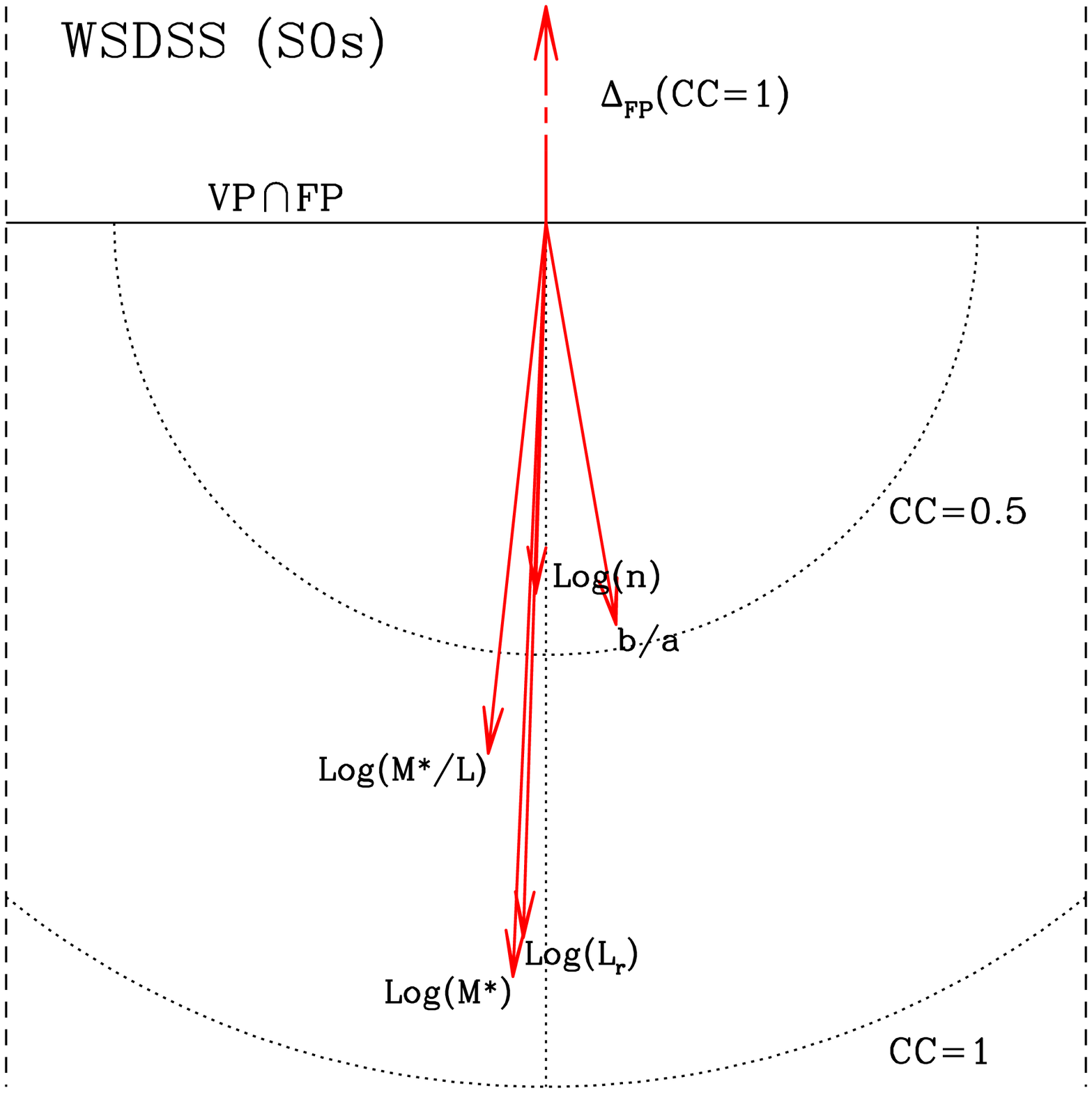}
\end{array}$
\vspace{-0.5truecm}
\caption{
   Similar to Fig~\ref{Fig4_FP}, but for the samples $ATLAS3D$ (upper panels)
  and $WSDSS$ (lower panels) samples.}
\label{Fig9_FP}
\end{center}
\end{figure*}

Figure~\ref{Fig9_FP} illustrates, for the samples $ATLAS3D$ and $WSDSS$
too, the results of the analysis already performed for the $WINGS$
sample about the directions of maximum correlation onto the FP for the
various physical parameters possibly involved in its {\it tilt} (see
Fig.~\ref{Fig4_FP}). Comparing Fig~\ref{Fig9_FP} with Fig.~\ref{Fig4_FP},
we note that: {\it (v)} the arrows relative to $M^*/L_V$ and $n$, for
all three samples ($WINGS$, $ATLAS3D$ and $WSDSS$), are substantially in
agreement with each other as far as both the angle and the orientation
onto the FP are concerned. The alignment with $\Delta_{FP}$ is pretty
good (although with opposite orientation) and the correlation is in
general stronger for Es than for S0s; {\it (vi)} for the axial ratio
(b/a) the agreement among the three samples is less good, but still
reasonable if one takes into account the different surface photometry
techniques used by the three surveys. In this case the strength of the
correlation is greater for S0 than for E galaxies and the alignment
with $\Delta_{FP}$ is usually poor.

Finally, from the upper panels of Figure~\ref{Fig9_FP} (just $ATLAS3D$
sample) we note that: {\it (vi)} the arrows relative to both
$M^*/M_{Tot}$ and $V_{rot}/\sigma_e$ turn out to be very well aligned
with $\Delta_{FP}$ for Es, while there is a substantial lack of
alignment for the S0 sample.

>From the previous points it is clear that the results illustrated in
Section~\ref{sec41} for the $WINGS$ galaxy sample are substantially
confirmed by the $ATLAS3D$ and $WSDSS$ galaxies, the main difference
being the weaker correlation we found between $\Delta_{FP}$ and
$M^*/L$ for the $WINGS$ galaxy sample with respect to the two
comparison samples. In general, however, all three data samples
support the conclusion that both the stellar populations
  ($M^*/L$) and the structural non-homology ($n$) are
responsible for the FP {\it tilt}.

As in the case of the $WINGS$ sample, also for the $ATLAS3D$ and
$WSDSS$ data, this conclusion is supported by the MVRA (see the
  first two rows of each sample in Table~\ref{MVRA1}).  The
conclusions are not so straightforward for the other tested quantities
(b/a, $V_{rot}/\sigma_e$ and $M^*/M_{Tot}$). In fact, for the
$ATLAS3D$ and $WSDSS$ samples, the PCA provides four (marginally five:
46.0\%, 22.3\%, 12.9\%, 9.2\%, 6.1\% and 3.5\%) and three (marginally
four: 50.8\%, 26.8\%, 14.6\% and 7.8\%) significant eigen-vectors,
respectively. The combined inspection of the upper plots in
  Fig.~\ref{Fig9_FP} and of Table~\ref{MVRA1}, seems to indicate that,
  of the two 'twin' quantities $b/a$ and $V_{rot}/\sigma_e$ (see
  Fig.~\ref{Fig8_FP}), the first one might be the most significant
  (\ie~ that driving the correlation with $\Delta_{FP}$) for the S0
  galaxies, while the last one might be the most significant for
  elliptical galaxies.  Finally, the the dark matter fraction too
  could be a not negligible ingredient of the FP {\it tilt}, at least
  for Es (see again the upper plots in Fig.~\ref{Fig9_FP}).

\begin{table}
\caption{Multi-variate regression analysis coefficients of the
relation: $\Delta_{FP}$=$\sum_i{c_i X_i}$+$const.$ for the 
$ATLAS3D$ and $WSDSS$ samples.}
\begin{tabular}{ccccc}\\
\hline\hline
   $X_i$             &   $c_i$  & $r.m.s.$ &  $t$ value & P($>\vert t\vert$) \\
\hline\hline
\multicolumn{5}{c}{$ATLAS3D$ sample}\\
\hline
$\log(M^*/L_V)$ & -0.439  &   0.064   &   -6.878   &   e-10   \\
$\log(n)$           & -0.456  &   0.059   &   -7.692   &   e-12      \\
$b/a$                & -0.026  &   0.063   &   -0.409   &   0.6830   \\
$-\log(V_{rot}/\sigma_e)$ & -0.119  &   0.037   &   -3.228   &   0.0016   \\
$-\log(M^*/M_{tot})$ & -0.788  &   0.181   &   -4.354   &   0.0001   \\
\hline
\multicolumn{5}{c}{$WSDSS$ sample}\\
\hline
$\log(M^*/L_V)$ & -0.921  &   0.042   &   -21.86   &   e-16   \\
$\log(n)$           & -0.341  &   0.028   &   -11.86   &   e-16      \\
$b/a$                & -0.064  &   0.020   &   -3.241   &   0.0012   \\
\hline\hline
\end{tabular}
\label{MVRA1}
\end{table}

\subsection{The amount of non-homology}\label{sec5}

The first attempt to estimate the non-homology term $K_V$ defined in
eq.~\ref{eqMtot} for a spherical galaxy can be traced back to
\cite{Poveda}, who derived values around $\sim3$. More recently,
\citet{Cappellari} compared the virial ($M/L \propto \re\sigma^2/L$)
and Schwarzschild estimates of $M/L$ obtaining $K_V=5.0\pm0.1$, while
\cite{Gallazzi} found values between 5 and $\sim 7$.

In the previous sub-sections we have found that the dark matter
fraction ($1-M^*/M_{Tot}$) and the structural and dynamical
non-homology ($n$, $\varepsilon$, $V_{rot}/\sigma$) play a role at
least as important as the stellar population effects ($M^*/L$) in
determining the bulk of the FP {\it tilt}. These results imply that the
term $K_V$ in the expression~(\ref{eqKV1}) cannot be constant. This is
at variance with the finding of \citet{Cappellari} and
\citet{Cappellari3}, who claimed that the mass-to-light-ratio has the
whole responsibility of the FP {\it tilt}, finding an almost constant
value of $K_V$ and ruling out the non-homology as a possible driver
of the {\it tilt} itself.

\begin{figure}
\begin{center}
\includegraphics[width=70mm,angle=-90]{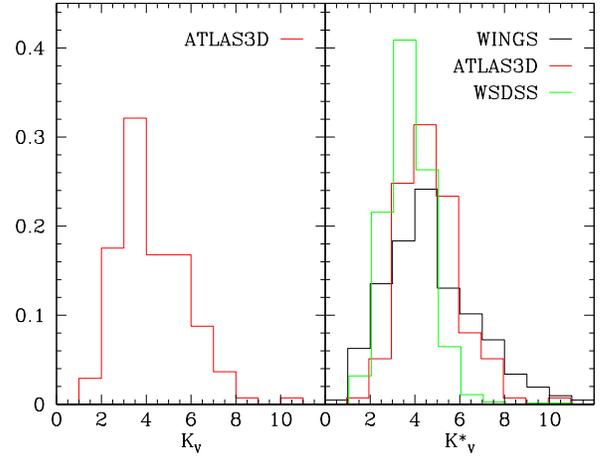}
\caption{
  {\bf Left panel}: histograms of the $K_V$ obtained from
  eq.~\ref{eqKV1} for the $ATLAS3D$ sample; {\bf Right panel}:
  histograms of the $K^*_V$ obtained from eq.~\ref{eqKV1}. The red and
  green histograms (top panel) refer to the $ATLAS3D$ and $WSDSS$ data,
  respectively, while the black histogram report the same histogram
  relative to the $WINGS$ sample.}
\label{Fig10_FP}
\end{center}
\end{figure}

Trying to clarify this point, we used the $ATLAS3D$ data to compute
$K^*_V$ from eq.~(\ref{eqKV1}) and $K_V$ from the ralation
$K_V=K^*_V*k_m$ (see Section~\ref{sec1}), where $k_m=M_{Tot}/M^*$.  We
also computed $K^*_V$ (again from eq.~\ref{eqKV1}) for both the
$WINGS$ and $WSDSS$ data. In the left panel of Figure~\ref{Fig10_FP} we
show the distribution of $K_V$ for the $ATLAS3D$ sample.

\begin{figure}
\begin{center}
\includegraphics[width=90mm]{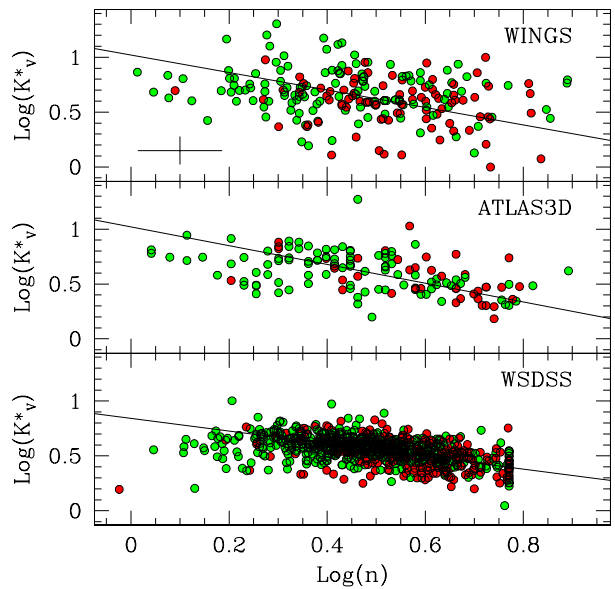}
\caption{
  The correlation between the $K^*_V$ and $\log(n)$
  for the $WINGS$, $ATLAS3D$ and $WSDSS$ samples. 
  Symbols are as in the previous figures.}
\label{Fig11_FP}
\end{center}
\end{figure}

The $K_V$ distribution of the $ATLAS3D$ sample looks at variance with
the very finding claimed by \citet{Cappellari} and \citet{Cappellari3}
about the small scatter of $K_V$ ($\sim 0.1$). We guess that, rather
than to the scatter, such a small value actually corresponds to the
estimated uncertainty found by \citet{Cappellari} for the best-fit
slope of the relation between the virial and Schwarzschild $M/L$
estimates (see Fig.13 therein).  In the right panel of
Figure~\ref{Fig10_FP} we compare the distributions of the $K^*_V$ obtained
for all three samples ($WINGS$, $ATLAS3D$ and $WSDSS$).  We remind that,
according to its very definition (see Sec.~\ref{sec1}), besides the
structural and dynamical non-homology, the $K^*_V$ term also
parametrizes the dark matter contribution (unknown, in the case of the
$WINGS$ and $WSDSS$ samples).  The $K^*_V$ distribution for the $ATLAS3D$
sample, where the dark matter contribution has been directly
estimated, turns out to be in fair agreement with (although narrower
then) that of the $WINGS$ sample and slightly shifted upwards with
respect to that of the $WSDSS$ sample. Actually, the three
distributions are not so different and, in any case, the statement
about the constancy of $K_V$ and $K^*_V$ does not seem to be supported
by the observations. This is also illustrated in Figure~\ref{Fig11_FP}, where
$K^*_V$ is shown to depend on the Sersic index in all three samples.

\subsection{The hybrid solution}

In this Section we have strongly promoted the so called `{\it hybrid
  solution}' of the FP problem. As already pointed out in the
introduction, the idea is actually not new at all. Several previous
works have also found that different physical mechanisms might concur
in shaping the FP \citep[{\it tilt} and thickness; see e.g.][among
others]{Ciotti,PrugSimien,Pahre}.  In their pioneering paper,
C96 first highlighted, from a theoretical point of view,
that a {\it fine-tuning} is required to explain the FP {\it tilt} with
single physical effects, such as \eg\ structural non-homology or DM
distribution. They concluded that: "...it remains the possibility of a
{\it hybrid} origin for the {\it tilt}, with more than one effect
contributing to tilting the FP, for example a small progression of
anisotropy, DM concentration, and shape $n$, coupled with a modest
increase of $M^*/L$."  \cite{PrugSimien} suggested that, adding
together the contributions of stellar population effects, rotational
support, and non-homology, one can fully account for the {\it tilt} of
the FP.  \cite{Pahre} presented a comprehensive model in which stellar
population gradients and systematic deviations of the internal
dynamical structure of ETGs from the homology, simultaneously explain
the differential {\it tilt} of the FP among different bandpasses, the
slope of the near-infrared FP and the slope of the $Mg_2-\sigma$
relation.

Our approach is in the groove of the {\it hybrid} solution, but
differs from the previous ones since we do not try to attribute the FP
properties to the mere addition of several contributors. Instead,
supported by our data, we believe that the various physical mechanisms
shaping the FP are mutually entangled. In our picture, according to
the $n - M^* - M^*/L$ relation found by \cite[][see
Section~\ref{sec7}]{Donofrio2}, at increasing the stellar mass, ETGs
become (on average) 'older' and more centrally concentrated. Such
twofold behaviour is due to the fact that the old stellar populations
tend to be more centrally concentrated than the young ones
(Fig.~\ref{Fig13_FP}). In the following sections, we show that,
besides to easily explain the differential FP tilt between the V- and
K-band, this natural {\it fine-tuning} mechanism might also drive the
main FP properties ({\it tilt} and thickness).

  To this concern, we note that it would be misleading to believe
  that, in the hyperspace defined by \re, \muem, $\sigma$, $M/L$, $n$,
  $\varepsilon$ (and possibly $V_{rot}/\sigma$ and DM fraction), the
  hybrid solution implies the existence of an hyperplane around which
  galaxies crowd with a scatter significantly smaller than that around
  the FP. First of all, if we assume that the scalar Virial Theorem is
  governing the galaxy structure and dynamics, the only hyperplane
  around which all galaxies are distributed with a scatter just due to
  measurement errors is that expressed by the eq.~\ref{eqvir5}. In
  this equation, besides the observables \re, \muem\ and $\sigma$, we
  have to deal with the unknown quantities $M/L$ and $K_V$. Since we
  adopt $n$, $\varepsilon$, $V_{rot}/\sigma$, $M^*/L$ and the dark
  matter fraction as proxies of $M/L$ and $K_V$, the fact that
  $\Delta_{FP}$ depends on these quantities just tells us that each
  galaxy contributes to the FP {\it tilt} through its own $M/L$ and
  $K_V$. Moreover, since the above proxies are only approximately
  correlated with $M/L$ and $K_V$, if we use them, together with \re,
  \muem and $\sigma$, to form an hyperspace, we introduce the
  additional (intrinsic) scatter of the correlations between $K_V$ and
  the quantities $n$, $\varepsilon$ and $V_{rot}/\sigma$, as well as
  that relative to the correlation between $M/L$ and $M^*/L$.  Thus,
  we expect that around the hyperplane the scatter is not smaller than
  that found around the FP. This unfailingly happens. In fact, the
  scatter of galaxies in our {\it Sample-II} around the hyperplane
  best fitting the quantities \re, \muem, $\sigma$, $M/L$, $n$ and
  $\varepsilon$ turns out to be $\sim 0.09$, quite similar to that
  found in Sec.~\ref{sec3} around the FP in both the V- and K-band.

\section{Origin of the differential tilt}\label{sec6}

Following the same logical thread of Sec.~\ref{sec4} and using our
$WINGS$ {\it Sample II}, we investigate in this section the origin of
the differential FP {\it tilt} observed between the $V$- and
$K$-band. In this case we parametrize the position of galaxies on the
FPs through the difference between the planes themselves
[eq.~(\ref{ourFP}) and (\ref{ourFPK})], averaged over the values of
$\muem$ in the two bands (the velocity dispersion being the same in
both equations).  In practice, we assume the analogous of
$\Delta_{FP}$ (see eq.~\ref{eqvirobs1}) to be, in this case, the
quantity:

\bigskip\noindent
$\Delta FP_{VK} = (a_V-a_K)\log(\sigma)+(b_V-b_K)(\langle\mu\rangle^V_e+\langle\mu\rangle^K_e)/2$,

\bigskip\noindent
where $a_V$,$b_V$,$\muem^V$ and $a_K$,$b_K$,$\muem^K$ are the FP
coefficients and average surface brightness in the V- and K-band,
respectively, while the term $(\muem^V+\muem^K)/2$ parametrizes the
band-averaged position of each galaxy along the \muem~axis.  Then, we
look whether such quantity correlates with some proxies of the stellar
population and luminosity structure.

\begin{figure}
\begin{center}
\vspace{-1truecm}
\includegraphics[width=95mm]{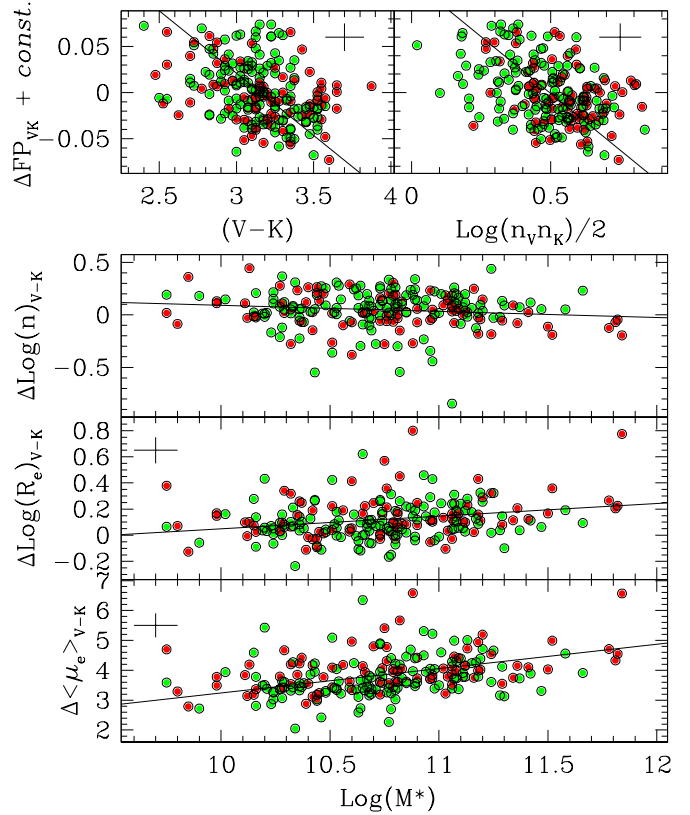}
\caption{
  {\bf Upper panels}: the difference between the FPs in the $V$- and
  $K$-band (see text for details) as a function of The $(V-K)$ color
  (left panel) and the band-averaged logarithm of the Sersic index
  (right panel) for the galaxies in the $WINGS$ sample; {\bf Lower
    panels}: the Sersic index (top panel), the effective radius
  (middle panel) and effective surface brightness (bottom panel)
  differences between the $V$- and $K$-band as a function of the
  stellar mass for the galaxies in the $WINGS$ sample. Symbols as in
  Fig.~\ref{Fig2_FP}.}
\label{Fig12_FP}
\end{center}
\end{figure}

\begin{figure}
\begin{center}
\includegraphics[width=90mm]{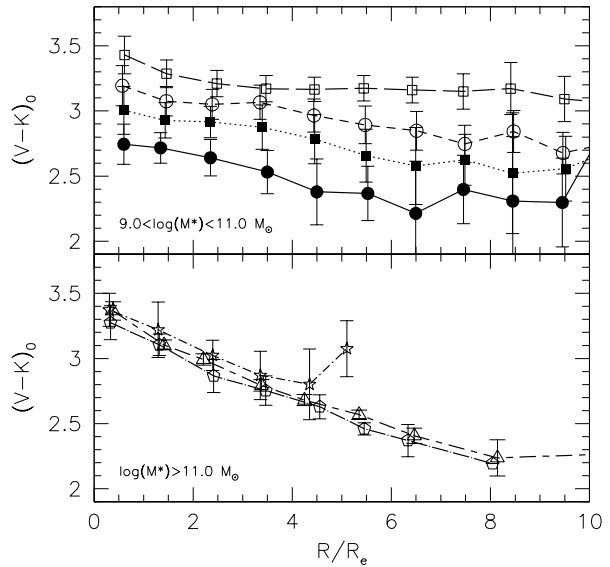}
\caption{
  {\bf Top panel:} the mean $V-K$ color profiles of our ETGs corrected
  for galactic extinction and K-corrections for different bin of mass:
  filled circles ($9.0<\log(M^*)<9.5$, 92 objects), filled squares
  ($9.5<\log(M^*)<10.0$, 199 objects), open circles
  ($10.0<\log(M^*)<10.5$, 250 objects), open squares
  ($10.5<\log(M^*)<11.0$, 200 objects). Each bin is the average of
  several growth curves measured by GASPHOT. {\bf Bottom panel:} the
  same for very massive galaxies: stars ($11.0<\log(M^*)<11.5$, 82
  objects), open pentagons ($11.5<\log(M^*)<12.0$, 18 objects), open
  triangles ($12.0<\log(M^*)<12.5$, 5 objects). }
\label{Fig13_FP}
\end{center}
\end{figure}

In the two topmost panels of Figure~\ref{Fig12_FP} the above correlations
are shown for the galaxies in our sample. In particular, the color
$V-K$ [$\simeq \Delta\log(M^*/L)_{V-K}$] and the band-averaged logarithm
of the Sersic index are used as proxies of the stellar population and
structure, respectively. In both cases the correlations are clear
($C.C.\sim 0.36$ and $0.38$, respectively, with high significances in
both cases), thus indicating that both the stellar population and the
luminosity structure contribute to the FP {\it tilt} difference
between the $V$- and $K$-band. How these factors operate in producing
such a result is clarified by the Figure~\ref{Fig13_FP} and by the three
bottom panels in Figure~\ref{Fig12_FP}.

Figure~\ref{Fig13_FP} shows the average color profiles of ETGs in our
sample for different stellar mass intervals. The color profiles are
derived from the GASPHOT growth curves, corrected for galactic
extinction and K-correction and binned as a function of $R/R_e$.
Besides confirming the well known inward reddening of ETGs, the
Figure~\ref{Fig13_FP} shows that their average color gradients are
quite shallow for galaxies in the low/intermediate mass regime
[$\log(M^*/M_{\odot})<11$], becoming relevant for more massive
objects. We do not investigate here the origin of such different
gradients (which of course can be related to metallicity effects and
to the different star formation histories of ETGs). We just note here
that our result is in good agreement with the finding of
\cite{LaBarb2}, which is based on a much larger sample \citep[but see
][for a different result]{Tortora2}. It suggests that the differential
FP {\it tilt} between the $V$- and $K$-band is originated by the
different behaviour of the average luminosity profiles for different
ranges of stellar mass in the two wavebands. In fact,
Figure~\ref{Fig13_FP} shows that, at increasing the stellar mass, the
old (red) stellar population tends to be more centrally concentrated,
most of the difference concerning massive galaxies
[$\log(M^*)>11$]. This color behaviour produces the trends observed in
the bottom panels of Figure~\ref{Fig12_FP}. In fact, it implies that,
at increasing the stellar mass, the luminosity profiles of ETGs appear
progressively steeper (inward) in the $K$- than in the $V$-band
[decreasing $\Delta\log n$ and increasing $\Delta\log R_e$ and
$\Delta\muem$]. Since the position of galaxies on the FP also
corresponds to a mass sequence (see upper-left panel of
Fig.~\ref{Fig2_FP}; see also Fig.~\ref{Fig4_FP}), the trends shown in
the bottom panels of Figure~\ref{Fig12_FP} produce, in turn, the
differential {\it tilt} between the $V$- and $K$-band. In some sense,
saying that the differential FP {\it tilt} between the V- and K-band
is due to the mass-dependent spatial distribution of stellar
populations inside ETGs (\ie~stellar population effects) is equivalent
to say that it is due to the different behaviour of the luminosity
profiles in the two bands as a function of stellar mass
(\ie~structural effects). This means that a sort of entanglement
between structural and stellar population effects is at the origin of
the differential {\it tilt} between the V- and K-band.  We note in
passing that, if we try to figure out the above reasoning in the
viewpoint of the usual $M/L$ vs. non-homology diatribe, the
differential FP {\it tilt} between the $V$- and $K$-band must be
ascribed to both the luminosity structure (different luminosity
profiles in the two bands: non-homology) and stellar population
($M/L$) effects, the two contributions being mutually entangled.

As already noted at the end of Sec.~\ref{sec41} for the results
illustrated therein, the findings in this Section and the related
analyses remain unchanged when computing $\Delta_{FP}$ from the
coefficients of eqs.~\ref{ourFP1} and \ref{ourFP1K} rather than those
of eqs.~\ref{ourFP} and \ref{ourFPK}, \ie when using the velocity
dispersion (instead of the effective radius) as independent variable
in the fit of the FP. {\bf However, similarly to what noted for $M^*/L$ in
Sec.~\ref{sec41}, the color $V-K$ turns out to be less strongly
correlated with $\Delta_{FP}$ than in the usual FP formulation
($CC\sim 0.28$ vs. $-0.36$), again suggesting that the influence of
the stellar populations on the FP {\it tilt} is rather uncertain.}

\section{The thickness of the FP}\label{sec7}

In Section~\ref{sec3} we have shown that the thickness of the FP
cannot be entirely due to observational errors. These can contribute
to nearly half of the observed scatter of $\sim0.09$. The additional
(intrinsic) scatter of the FP implies that ETGs do not deviate from
the plane more than $\sim 15\%$. As mentioned in the introduction,
this scatter has been mainly attributed to stellar population effects
through the age of galaxies.

\begin{figure}
\begin{center}
\includegraphics[width=90mm]{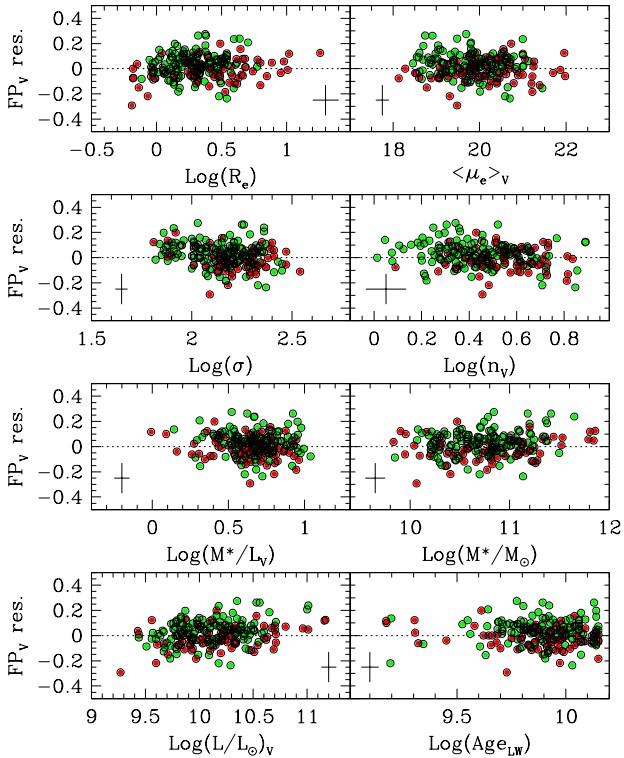}
\caption{
  Residuals of the best-fit FP in the $V$-band from $WINGS$ data
  (eq.~\ref{ourFP}) vs. mass-to-light ratio (upper-left panel),
  stellar mass (upper-right panel), $V$-band luminosity (bottom-left
  panel) and luminosity-weighted age (bottom-right panel) of 
  galaxies in our sample.}
\label{Fig14_FP}
\end{center}
\end{figure}

Figure~\ref{Fig14_FP} shows the residuals of the best-fit FP in the
$V$-band (eq.~\ref{ourFP}) for the galaxies in our {\it Sample II} as
a function of some interesting quantities.  No significant trends are
found, in particular with the luminosity weighted Age. The weak
correlations with $(M^*/M_\odot)$ and $(L/L_\odot)_V$ is likely
indicating that the main driver of the FP residuals is the selection
effect connected with the data sample used (\ie\ with the range of
stellar masses and total luminosities). This would confirm the
suggestion given in Paper-I that the FP is likely a bent surface and
that a careful choice of the data sample is mandatory before drawing
any conclusion about the FP properties (including thickness; see also
Section~\ref{sec8}).

In lack of any clear indication coming from the analysis of the
residuals, we may wonder about whether the observed scatter around the
FP is consistent with our finding that the FP {\it tilt} is originated
by the simultaneous influence of different physical factors
(non-homology, mass-to-light-ratio, dark matter fraction).

According to C96, even if the range of Sersic indices spanned by ETGs
is in principle able to explain by itself the observed FP {\it tilt},
the small scatter around the plane would require in this case an {\it
  'ad-hoc'} fine-tuning. Using dynamical, isotropic models of
spherical galaxies with Sersic profiles and without DM, C96 found that
the observed FP tightness would require a tight correlation between
the Sersic index ($n$) and the $k_1$ parameter of the $k$-space
defined by \cite{BBF} [$k_1=(2\log\sigma+\log\re)/\sqrt{2}$]. In
particular, if the FP {\it tilt} is entirely due to Sersic index
variation, the maximum allowed range of $n$ at any given $k_1$ should
not exceed $\sim 1.0$ (Figure~5 therein). Instead, using the small
dataset of Virgo galaxies studied by \cite{Caon}, C96 claim that the
observed scatter of the $n$-$k_1$ relation turns out to be nearly 3
times larger (Figure~6 therein).

Here, using the ETGs in our sample, we show that the intrinsic
scatter of the $n-k_1$ relation may be consistent with the required
theoretical strip defined by C96.

\begin{figure}
\begin{center}
\includegraphics[width=70mm, angle=-90]{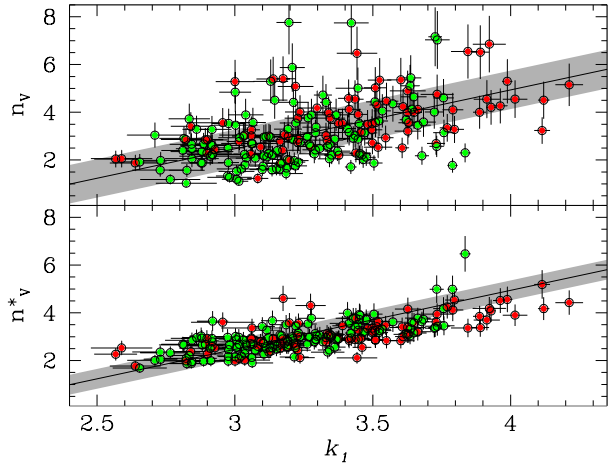}
\caption{
  {\bf Upper panel}: the relation between the observed Sersic index
  $n_V$ and the k-space parameter $k_1$. The black solid line marks
  the linear best fit, and the gray strip shows the intrinsic $r.m.s.$
  uncertainty once taken into account the uncertainties in both
  coordinates; {\bf Lower panel}: same relation of the upper panel,
  but replacing the observed value of the Sersic index ($n_V$) with
  the one ($n^*_V$) obtained through eq.~(\ref{eqfp} he values of $n$
  derived from eq.~\ref{eqfp}. The errors on $n^*_V$ are obtained by
  propagating the uncertainties of $M^*$ and $M^*/L_V$.}
\label{Fig15_FP}
\end{center}
\end{figure}

The upper panel of Figure~\ref{Fig15_FP} shows the relationship between
$n$ and $k_1$ for the $V$-band. Once the
uncertainties on both coordinates ($n_V$ and $k_1$) are taken into
account, the intrinsic scatter of the relation (gray strip in the
figure) turns out to be: $2\times r.m.s.\sim$1.58.  This means that
the intrinsic variation of $n_V$ at each $k_1$ is just a factor $\sim
1.5$ larger than the theoretical strip. Taking into account that such
strip is obtained assuming spherical, isotropic models without DM and
that the structural non-homology is likely not the only contributor to
the FP {\it tilt} (as assumed by C96), we can guess that for
our galaxy sample the {\it intrinsic} range of variation of $n_V$ at
each $k_1$ might be comparable to the theoretical one ($\sim 1$).
Therefore, the {\it 'ad hoc'} fine-tuning required by models seems to be
actually in place and we are lead to conclude that even the structural
non-homology alone might simultaneously explain the {\it tilt} and the tightness of
the FP.

Trying to identify the origin of such fine-tuning, we recall that
\citet{Donofrio2} discovered a bi-variate correlation linking $n$,
$M^*$ and $M^*/L$ (nMML relation, hereafter). The best-fit of this
relation for our galaxy {\it Sample II} in the V-band turns out to be:

\begin{equation}\label{eqfp}
\log(n^*_V)=0.285*log(M^*)-0.617*\log(M^*/L_V)+const.
\end{equation}

\noindent
with $r.m.s. = 0.14$. 

The existence of a relationship between structure, mass and stellar
populations in ETGs has been claimed in various ways in the
literature.  \cite{Valentinuzzi2} \citep[see also][]{Poggianti} found that both
in clusters and in the field, at a given galaxy size, more massive
objects have older luminosity weighted ages, while at a given mass,
larger galaxies have younger luminosity weighted ages, \ie\ there is a
correlation between compactness and age of ETGs.  More recently,
\cite{Wuyts} confirmed that objects with different Sersic index
populate in a different way the SFR-Mass relation, built for the whole
Hubble sequence (their Figure 1).

\begin{figure}
\begin{center}
\includegraphics[width=80mm]{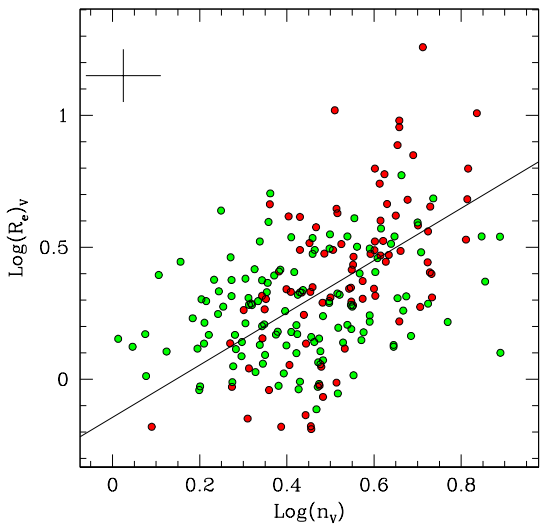}
\caption{
  The effective radius \re\ as a function of $n_V$ in log units. The
  full line shows the linear best-fit and symbols are as in the
  previous figures.}
\label{Fig16_FP}
\end{center}
\end{figure}

It is worth stressing that the direction of the nMML relation is in
agreement with the findings of \citet{Valentinuzzi2} and
\cite{Poggianti}. In fact, given the well known direct correlation
between the Sersic index and the effective radius \citep[see
Fig.~\ref{Fig16_FP}; see also][Fig.5 therein]{Caon}, the
equation~(\ref{eqfp}) confirms the above mentioned link between age
and compactness.  Therefore, the nMML relation coexists with the FP
relation and could actually represent the conspiracy between stellar
population and structure (fine-tuning) which, according to C96, is
required to explain the FP tightness if non-homology is responsible of
the FP {\it tilt}.

As an exercise, we plot in the lower panel of Figure~\ref{Fig15_FP} the
same relation shown in the upper panel, but replacing the observed
value of the Sersic index ($n_V$) with the one ($n^*_V$) obtained
through eq.~(\ref{eqfp}). The uncertainties in this figure are
computed by error propagation. It is interesting to note that the
intrinsic scatter of the new relation $n^*_V$-$k_1$ ($r.m.s. \sim
0.50$) turns out to be fully consistent with the above mentioned,
'theoretical' one.  It is also worth stressing that this very tight
relation involves four largely independent quantities, \ie\ \re\ and
$\sigma$ for the abscissa, $M^*$ and $L_V$ for the ordinate.

Finally, we note that the nMML relation is someway implicit in the
above eq.~(\ref{eqKV1}).  In fact, using the correlation

\begin{equation}
\log(K^*_V) = -1.35\log(n)+0.634
\end{equation}

\noindent
shown in the uppermost panel of Figure~\ref{Fig11_FP} between $n$ and $K^*_V$
(the last one computed from eq.~\ref{eqKV1}) and the correlation
between $\log(M^*)$ and $\Delta_{FP}$ shown in the upper-left panel of
Figure~\ref{Fig2_FP} (see the corresponding linear coefficients in
Table~\ref{Tab0}), the equation~(\ref{eqKV1}) takes the form:

\begin{equation}\label{eqfp1}
\log(n^*_V)=0.274*log(M^*)-0.740*\log(M^*/L_V)+const. 
\end{equation}

\noindent
which, given the large fitting uncertainties, turns out to be
remarkably similar to eq.~(\ref{eqfp}). Thus, the nMML relation seems
to simultaneously explain the {\it tilt} and the tightness of the
FP. In some sense, the very existence of the FP {\it tilt} seems to
imply the FP tightness, \ie\ the two things appear to be entangled
through the nMML relation.

\section{The mass-to-light ratio of high redshift galaxies}\label{sec8}

Since structure and stellar population of ETGs seem to be linked each
other, the common practice of deriving the total mass-to-light ratio
($M_{tot}/L$) of ETGs by means of the FP assuming the perfect homology
($K_V=constant$) might be not correct, in particular for high redshift
galaxies.  It is well known that the FP coefficients are believed to
vary with redshift, both in clusters \citep[see \eg\
][]{vanDokkum,Sperello,Jorgensen} and in the field \citep{Treu05}. If
one uses the FP relation to derive the dynamical $M_{tot}/L$ assuming
the homology, the above variation implies a progressive change of the
mass-to-light ratio with redshift: $d\log(M_{tot}/L_B)/dz \sim -0.7$
\citep[see e.g.,][]{Treu05}.  This behavior has been commonly
explained invoking the ``downsizing'' phenomenon (see \eg\
\citealt{Sperello,Jorgensen}), \ie\ the idea that the mass-to-light
ratio evolves at different rates for low and high mass galaxies.
Actually, while for massive galaxies the FP appears nearly unchanged
since $z\sim1$, for low-mass galaxies a progressive displacement from
the local FP is observed at increasing redshift
\citep{vanderWel,Treu05}.

In this section we indicate two warnings about this kind of
reasoning. The first one is that the existence of the nMML relation
implies that, besides the stellar populations, the galaxy structure
too varies with redshift \citep[see e.g.,][]{Chevance}. The second one
is that a Malmquist bias might mimic the ``downsizing''
phenomenon. This should advise people to carefully take into account
both effects before drawing any conclusion about the evolution of the
mass-to-light ratio.

In order to illustrate the above mentioned second warning,
we first derived the FP coefficients using the high redshift data of
\cite{Sperello}, obtained from the K20 survey at redshift $z\sim1$
\citep{Cimatti}. We choose these data for our analysis because all the
values of the Sersic index $n$, the dynamical masses, and the total
luminosities in the $B$ band rest frame of the galaxies are given for
this sample, allowing a direct comparison with the $WINGS$ data.
The fit of the FP with our MIST procedure provides the following
coefficients for the high redshift dataset:

\begin{equation}
\log(\re)=0.74\log(\sigma)-0.81\log(\langle I\rangle_e)+3.85
\end{equation}

\noindent
with errors in the coefficients $\pm0.13$, $\pm0.04$ and $\pm0.34$,
respectively. In this formulation \re\ is given in $pc$ and $\langle
I\rangle_e$ in $L_{\odot}/pc^2$.  These values of the FP coefficients
are very close to those obtained by \citet{Sperello}.  Comparing these
coefficients with those observed in the $B$ band rest-frame for Coma
by \cite{Jorg96}, the authors concluded that the strong {\it tilt}
difference between low and high redshift FPs implies the existence of
a ``downsizing'' mechanism, \ie\ a different time-scale variation of
the dynamical $M/L$ ratio for low and high mass galaxies. In addition,
they showed that the $M/L - M$ relation derived from the FP at high
redshift is quite different from that derived using a low redshift
dataset of ETGs.

In Paper-I, based on a much larger galaxy sample, we found that the FP
is likely a bent surface (see also \citealt{Desroches}) and that this
produces a systematic variation of the FP coefficients at varying the
bright-end cut-off of the galaxy sample.

Trying to check if this effect could result in a sort of {\it
  Malmquist bias} when moving at high redshift, we have obtained, for
each galaxy in our sample, a rough estimate of the dynamical
mass $M_{tot}$ from the virial eq.~(\ref{eqMtot}) using the expression
of $K_V$ (as a function of $n$) proposed by \cite{Bertin} for
one-component, spherical, non-rotating, isotropic \rnn\ stellar
systems:

\begin{center}
\begin{equation}\label{eq3}
K_V(n)=\frac{73.32}{10.465+(n-0.94)^2}+0.954.
\end{equation}
\end{center}

\begin{figure}
\begin{center}
\includegraphics[width=90mm]{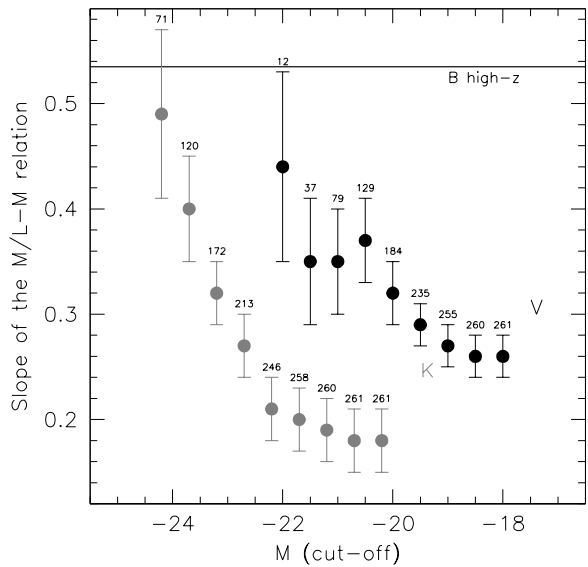}
\caption{The slopes of the $M_{tot}/L - M_{tot}$ relation, obtained for
  different cuts in absolute magnitudes in the $V$ (black dots) and
  $K$ (grey dots) bands. The number of galaxies used for each fit
  are reported close to the dots. The horizontal line corresponds to
  the slope of the high redshift $M_{tot}/L_B-M_{tot}$ relation derived from the
  data of \citet{Sperello}}
\label{Fig17_FP}
\end{center}
\end{figure}

Figure~\ref{Fig17_FP} reports the slopes of the best-fit regressions
$M_{tot}/L-M_{tot}$ obtained for different cut-off magnitudes
from our galaxy sample in the $V$- and the $K$-bands (black and
grey dots, respectively).

This figure shows that the slope of the $M_{tot}/L-M_{tot}$ relation
in the $V$ and $K$ bands increases when the cut-off luminosity
increases (as it occurs for high redshift data sample that
progressively loose faint objects). The same result is obtained when
the mass-to-light ratio is derived from the combination of the FP and
VP equations (eq.~\ref{fp} and \ref{eqvir5}, respectively), \ie\
taking into account the variation of the FP coefficients with the
cut-off luminosity.  Using the high redshift data provided by
\citet[][Table~2 therein]{Sperello}, we derived a slope of 0.51 for
the $M_{tot}/L_B - M_{tot}$ relation at high redshift (see the
horizontal line in Figure~\ref{Fig17_FP}). Note that Figure~8 in
\cite{Sperello} gives a cut-off magnitude for their high redshift
sample at $M_B\sim-20.1$, which roughly corresponds to $M_V\sim-21$
and $M_K\sim-24.2$ (assuming the typical colors of ETGs: $B-V\sim1$
and $B-K\sim4.2$). It follows that the trends shown in
Figure~\ref{Fig17_FP} could be (at least in part) due to the observed
change in the FP coefficients and in the slopes of the $M/L - M$
relation that occur when the sample is progressively cut by simulating
high redshift observations.

We are aware that many works have shown in different ways that M/L
evolves differently as a function of mass (see above
references). Here, we just wish to warn that for high redshift samples
the selection effects must be carefully taken into account before
attributing entirely to the ``downsizing'' phenomenon the changes of
the $M/L -M$ relation with redshift.

\section{Discussion and conclusions}\label{sec9}

Our analysis strongly suggests that the FP properties can be explained
only within an ``hybrid'' framework, in which several physical
mechanisms are at work together. They are: {\it (i)} the non homologous
structure and dynamics of ETGs; {\it (ii)} the stellar population
variation along the sequence of galaxy masses; {\it (iii)} the
increasing DM fraction within \re\ at increasing the galaxy mass.

In this work we have shown that the conspiracy between stellar
population and galaxy structure, which has been invoked by C96 to
explain the FP properties with just non-homology, does actually exist
and takes the form of the nMML relation. According to this relation,
for a given stellar mass, galaxies with high (low) Sersic index
values, have high (low) values of $M^*/L$. The coupling of structure
and the stellar population variations along the FP seems to be an
important physical mechanism behind its properties, since it turns out
to simultaneously explain the {\it tilt} and the scatter of the FP.
The data also suggest that the above mentioned conspiracy between
structure and stellar population is not the only mechanism at work. In
fact, the DM fraction seems also to play a role in shaping the FP,
although both the sign and the strength of the correlation with
$\Delta_{FP}$ turn out to be different for Es and S0 galaxies (see
Fig.~\ref{Fig9_FP} and Tab.~\ref{Tabatlas}). This again support the
``hybrid'' scenario to explain the FP properties.

Although the recent huge progresses of numerical simulations have
produced massive ETGs resembling the real ones and obeying the most
important scaling relations, at present we are not aware of
theoretical models of galaxy formation and evolution able to
simultaneously explain the FP, the nMML and the DM variation with
stellar mass.

Some interesting numerical simulations have shown that a series of
minor merging events (either dry or partially wet), could modify the
size of the galaxies and form extended stellar envelopes, while the
stellar density and the velocity dispersion decreases with time
\citep[see \eg][]{Nipotietal2009, Bezanson, Naab, Hopkinsetal2009,
  Oser}. This is the so called inside-out mechanism of galaxy
formation, predicting that the oldest and most massive galaxies have
very small sizes and are more compact than present day
objects. Several observations seem to confirm such hypothesis
\citep[see \eg]{Daddietal2005, Trujilloetal2006, vanDokkumetal2008,
  Cimattietal2008, vanderWeletal2008, Damjanovetal2009, Chevance},
even if other works found that the size evolution of individual
galaxies is modest when the age-size-mass relation is taken into
account \citep{Valentinuzzi1, Poggianti, Saracco}.  Unfortunately, the
minor merging scenario largely fails to reproduce the numerical
evolution of ETGs, given that the number density of ETGs turns out to
be $\sim$25 times larger today than at z$\sim$2.5, thus preventing the
size evolution of individual galaxies to be responsible of the claimed
mass-size evolution.

The semi-analytical models of hierarchical formation \citep[see
\eg,][]{DeLuciaetal2006, Almeidaetal2007, Gonzalezetal2009,
  Parryetal2009, DeLuciaetal2011} have also successfully explained
several observed features of ETGs, but fail in others. In particular
they overestimate the number of faint objects, under predict the sizes
of bright ETGs, and fail in reproducing the $\alpha$-enhancement that
increases with the mass of the galaxies. Moreover, it is not clear how
such models could be connected with the observed nMML relation.

A primordial activity of merging at $z \ge 2$ is not excluded by
\cite{MerlinChiosi2006, MerlinChiosi2007}, who suggested that a series
of merging of small stellar subunits could lead to the formation of
massive objects resembling the real ones in their morphology, density
profiles, and metallicity. In their scheme, a first generation of star
clumps inside primordial haloes of DM enriched the medium in metals
and made up the building blocks of larger systems with masses up to
$10^{12} M_{\odot}$. The hydro-dynamical simulations of
\citet{Merlinetal2012} include radiative cooling, star formation,
stellar energy feedback, re-ionization, and chemical enrichment. They
reproduce many observed features of ETGs, such as the mass-density
profiles, the mass-radius relation, the mass-metallicity relation,
etc. In their scheme objects with the same initial mass might have
experienced different star formation histories depending on the
initial halo over-density: the deeper the perturbation, the more
peaked, early and intense the star forming activity. This kind of
behavior could provide a viable explanation for the existence of the
nMML relation.

According to \cite{Nipoti}, the dissipationless collapse of initially
cold stellar distributions in preexisting DM haloes had an important
role in determining the observed weak homology of ETGs. The
end-products of their N-body simulations of collapses inside DM halos
have significant structural non-homology, with the Sersic index
spanning the interval $1.9 \leq n \leq 12$. Remarkably, their
parameter $n$ correlates with the DM fraction present within \re,
being smaller for larger dark-to-visible mass ratios. Unfortunately,
such models do not follow the star formation activity, so we do not
know if they are consistent with the nMML relation.

\citet{Hopkins} found that also the dissipational collapse following
the major merging of gas rich spirals could be usefully employed to
interpret the FP properties, the size-mass and velocity
dispersion-mass correlations. Their results, however, do not assign to
non-homology a key role in {\it tilting} the FP, contrary to our finding. It
is also not clear whether the major merging hypothesis, coupled with
the dissipational collapse, might be in agreement with the nMML
relation.

We actually believe that still much theoretical work should be done
before the problem of the formation of ETGs will find a robust and
definitive solution within the ``hybrid'' scenario strongly supported
by our analysis, whose main conclusions are the following:

\begin{itemize}

\item the {\it tilt} of the FP in the $K$-band is significantly
  smaller than in the $V$-band, but still substantial;
\item other than to the change of the stellar populations with
  galaxy mass, the bulk of the {\it tilt} in both wavebands can be
  attributed to a systematic variation of the structure of ETGs as a
  function of mass. This is proved by the observed dependence of the
  quantity $\Delta_{FP}$ (difference between the FP and a reference
  VP) on the Sersic index $n$ and on the axis ratios $b/a$ of the
  galaxies. This ``hybrid'' interpretation of the FP {\it tilt}, based
  on our WINGS galaxy sample, is also confirmed by the analysis of the
  $ATLAS3D$ and $WSDSS$ samples, for which the influence of the stellar
  mass-to-light-ratio on the FP {\it tilt} turns out to be greater than
  for the $WINGS$ sample;
\item using the data from the $ATLAS3D$ project and at variance with
  their claims, we find that again the ``hybrid'' interpretation of
  the FP should be preferred, since dynamical non-homology and DM
  effects seems also to play a significant role in producing the {\it
    tilt};
\item both the Principal Component and the Multi-Variate Regression
  Analyses suggest that most of the previously mentioned physical
  factors significantly contribute to the FP {\it tilt}. Again, this
  points towards an ``hybrid'' solution of the FP problem;
\item the differential {\it tilt} in the $V$- and $K$-band should be
  ascribed to the entangled variation of structure and stellar
  population between the two bands. This variation reflects the mass
  dependent color gradients ($V-K$) likely originated by metallicity
  effects;
\item the {\it tilt} and the small scatter around the FP are likely
  originated by the conspiracy between structure and stellar
  population, through the relation $\log(M^*) - \log(n) - \log(M*/L)$
  that works as a fine-tuning mechanism preserving the tightness of
  the FP plane along the whole sequence of masses. According to this
  relation, at increasing the stellar mass, ETGs become (on average)
  'older' and more centrally concentrated. Such twofold behaviour is
  due to the fact that the old stellar populations tend to be more
  centrally concentrated than the young ones, this fact being also
  responsible of the differential FP tilt between the V- and K-band;
\item the different slope of the dynamical $M/L - M$ relation at low
  and high redshifts must be used with care to derive the amount of the
  ``downsizing'' mechanism, since selection effects acting on the
  galaxy samples (\ie\ the Malmquist bias) mimic a differential rate
  of the $M/L$ evolution for small and massive ETGs.
\end{itemize}

\section*{Acknowledgments}

We warmly thank the anonymous referee for his comments and suggestions that have greatly
helped us to improve the paper.  We also acknowledge the partial financial support by contract INAF/PRIN 2011 ÒGalaxy Evolution with the VLT Survey Tele- scope (VST)Ó and PRIN-MIUR 2009
"Nature and evolution of superdense galaxies". Credits are finally due to the $SDSS$ and $ATLAS3D$ collaborations for
the data used in this work.

\label{lastpage}
\end{document}